\documentclass[useAMS,usenatbib]{mn2e}
\usepackage{graphicx}
\usepackage{times}
\usepackage{epsfig}
\usepackage{rotating}
\usepackage{sidecap}
\usepackage{longtable}
\usepackage{lscape}
\def\fullsrc{IGR J17480-2446}
\def\src{J17480}

\def\ltsima{$\; \buildrel < \over \sim \;$}
\def\simlt{\lower.5ex\hbox{\ltsima}}
\def\gtsima{$\; \buildrel > \over \sim \;$}
\def\simgt{\lower.5ex\hbox{\gtsima}}

\title[Pulse profile and spin evolution of {\fullsrc}]{The pulse
  profile and spin evolution of the accreting pulsar in Terzan 5,
  {\fullsrc}, during its 2010 outburst} \author[A. Papitto et
  al.]{A. Papitto$^{1,2}$\thanks{E-mail: papitto@ice.csic.es}, T. Di
  Salvo$^{3}$, L. Burderi$^{1}$, T. M. Belloni$^{4}$, L. Stella$^{5}$,
  E. Bozzo$^{6}$, A. D'A\`i$^{3}$, \newauthor C.Ferrigno$^{6}$,
  R. Iaria$^{3}$, S. Motta$^{4}$, A. Riggio$^{7}$,
  A. Tramacere$^{6}$\\ $^{1}$Dipartimento di Fisica, Universit\`a di
  Cagliari, SP Monserrato-Sestu, Km 0.7, 09042 Monserrato,
  Italy\\ $^{2}$Institut de Ci\`encies de l'Espai (IEEC-CSIC), Campus
  UAB, Fac. de Ci\`encies, Torre C5, parell, 2a planta, 08193
  Barcelona,  Spain \\ $^{3}$Dipartimento di Fisica, Universit\`a di
  Palermo, via Archirafi 36, 90123 Palermo, Italy \\ $^{4}$INAF -
  Osservatorio Astronomico di Brera, via E. Bianchi 46, 23807 Merate,
  Italy \\ $^{5}$INAF Osservatorio Astronomico di Roma, via Frascati
  33, 00040 Monteporzio Catone, Italy \\ $^{6}$ ISDC Science Data
  Center for Astrophysics of the University of Geneva, chemin
  d’\'Ecogia 16, 1290 Versoix, Switzerland \\ $^{7}$ INAF –
  Osservatorio Astronomico di Cagliari, Poggio dei Pini, Strada 54,
  09012 Capoterra, Italy}
\begin{document}

\pagerange{\pageref{firstpage}--\pageref{lastpage}} \pubyear{2011}

\maketitle

\label{firstpage}

\begin{abstract}

We analyse the spectral and pulse properties of the 11 Hz transient
accreting pulsar, {\fullsrc}, in the globular cluster Terzan 5,
considering all the available {\it Rossi X-ray Timing Explorer}, {\it
  Swift} and {\it INTEGRAL} observations performed during the outburst
shown between October and November, 2010.

By measuring the pulse phase evolution we conclude that the NS spun up
at an average rate of $<\dot{\nu}>=1.48(2)\times10^{-12}$ Hz s$^{-1}$,
compatible with the accretion of the Keplerian angular momentum of
matter at the inner disc boundary. This confirms the trend previously
observed by \citet{Pap11}, who considered only the first few weeks of
the outburst. Similar to other accreting pulsars, the stability of the
pulse phases determined by using the second harmonic component is
higher than that of the phases based on the fundamental
frequency. Under the assumption that the second harmonic is a good
tracer of the neutron star spin frequency, we successfully model its
evolution in terms of a luminosity dependent accretion torque. If the
NS accretes the specific Keplerian angular momentum of the in-flowing
matter, we estimate the inner disc radius to lie between 47 and 93 km
when the luminosity attains its peak value.  Smaller values are
obtained if the interaction between the magnetic field lines and the
plasma in the disc is considered.

The phase-averaged spectrum is described by thermal Comptonization of
photons with energy of $\approx 1$ keV. A hard to soft state
transition is observed during the outburst rise. The Comptonized
spectrum evolves from a Comptonizing cloud at an electron temperature
of $\approx$ 20 keV towards an optically denser cloud at $kT_e\approx$
3 keV. At the same time, the pulse amplitude decreases from $27\%$ to
few per cent, as already noted by \citet{Pap11}, and becomes
strongly energy dependent. We discuss various possibilities to explain
such a behaviour, proposing that at large accretion luminosities a
significant fraction of the in-falling matter is not channelled
towards the magnetic poles, but rather accretes more evenly onto the
NS surface.

\end{abstract}

\begin{keywords}
accretion, accretion discs -- stars: neutron -- X-rays: binaries -- pulsars: individual: {\fullsrc}

\end{keywords}

\section{Introduction}

The study of the frequency variations of the coherent signal emitted
by an accreting pulsar is one of the foremost techniques to study the
dynamical interaction between the rotating neutron star (NS in the
following) and the in-falling matter.  This is particularly the case
of disc-fed pulsars since the torque acting on the NS is expected to
reflect the specific angular momentum of the disc plasma before
it is captured by the pulsar magnetosphere, as well as the interaction
between the disc plasma and those magnetic field lines that close
through it \citep[see, e.g.,][and references therein]{Ghs07}.

The process of disc angular momentum accretion is also established as
the driver to accelerate a NS until it reaches a spin period of
few ms. According to the recycling scenario \citep[see,
  e.g.,][]{BhtvdH91} rotation-powered millisecond pulsars are spun up
by a long-lasting phase of accretion of mass and of the relative
angular momentum by a NS in a low-mass X-ray binary (LMXB). The
discovery of coherent ms pulsations in the X-ray emission of a
transient LMXB \citep{WijvdK98} provided a fundamental confirmation of
this scenario, also opening the possibility of studying the frequency
evolution of quickly rotating NS during the accretion phase.  However,
so far measuring the NS spin evolution while accreting of the known 14
accreting millisecond pulsars (AMSP) has been made often
  difficult by the intrinsic small torques exerted on the NS of these
systems. This is partly due to the relatively short timescales of the
accretion episodes, with outbursts lasting generally less than few
weeks, and to the relatively small peak outbursts luminosities, which
are not observed to exceed a level of few $\times10^{36}$ erg
s$^{-1}$.  To complicate further the task, timing noise was observed
to affect the phase of the pulse profile \citep[see,
  e.g.,][]{Brd06,Hrt09}. A spin-up at a rate of the order of that
expected according to standard accretion theories could then be
confirmed only in a couple of cases \citep{Fal05,Brd07,Pap08}.

Here we present the case of the 11 Hz accreting pulsar, {\fullsrc},
belonging to the globular cluster Terzan 5. The temporal evolution of
the 11 Hz spin frequency during a subset of all the observations
performed by the {\it Rossi X-Ray Timing Explorer} ({\it RXTE}) and
{\it Swift} was studied by \citet[][P11 in the following]{Pap11}, who
found an average spin up rate compatible with accretion of the
Keplerian angular momentum of disc matter. The analysis of the Doppler
shifts of the signal frequency allowed to measure the 21.27 hr orbital
period and to constrain the mass of the companion between 0.4 and 1.5
$M_{\odot}$ (see P11, and the references therein).  A candidate
  IR counterpart was proposed by \citet{testa2011} thanks to an
  observation performed by \textsl{ESO-VLT/NAOS-CONICA} while the
  source was in outburst. This candidate was then identified by
  \citet{patruno2012} in a \textsl{HST} archival observation performed
  when the source was in quiescence. From the presence of pulsations
at the minimum and maximum luminosity P11 estimated the surface
magnetic field of the NS to lie between a few times $10^8$ and a few
times $10^{10}$ G. These properties make this recently detected
mildly-recycled pulsar, the only discovered so far in a LMXB to spin
at a frequency between 10 and 100 Hz, a unique case study to directly
observe the effects of disc torque on the NS through the measure of
its spin evolution while accreting.  In this paper, we analyse in
detail the properties of the energy spectrum (Sec.~\ref{sec:spectrum})
and of the pulse profile (Sec.~\ref{sec:pulse}) of {\src} using all
the available X-ray observations performed by {\it RXTE}, {\it Swift},
and {\it INTEGRAL}, focusing on the effects of the torque exerted on
the NS by the disc accreting matter (Sec.~\ref{sec:timing} and
\ref{sec:torque}). The results obtained are then discussed in
Sec.~\ref{sec:discussion}.

\section{Observations}

In this paper we consider all the observations of {\fullsrc} ({\src}
in the following) performed by {\it RXTE}, {\it Swift} and {\it INTEGRAL}
during the intensive observational campaign which followed the
discovery of the source on 2010 October 10.365 (\citealt{Brd10}; MJD
55479.365, all the epochs are given in UTC). The monitoring of the
source ended at MJD 55519.130, due to a solar occultation of the sky
region where this transient lies. Such obstruction ceased only $\sim$
2 months later (MJD 55584.562); since further {\it RXTE} observations
failed to detect an excess above the background it has to be concluded
that the {\src} outburst ended during the period of non-visibility.

During the outburst the source showed a large number of X-ray bursts,
whose spectral and temporal properties are compatible with
thermonuclear burning of a mixed H/He layer on the NS surface
\citep{Mtt11,Ckr11,Lnr11}. Since we are here interested in the
analysis of the source properties during the non-bursting emission, we
discard $\sim$10 s prior the onset of each of the bursts that could be
significantly detected above the continuum emission, and a variable
interval of length between 50 and 160 s after, depending on the burst
decay timescale. During the analysis of observations taken in modes
with low temporal resolution (e.g. Standard 2 modes of PCA aboard {\it
  RXTE}) the length of the time intervals excluded from the analysis
was increased in order to equal the time-binning.

\subsection{RXTE}

In this paper we consider data obtained by the Proportional Counter
Array \citep[PCA; 2--60 keV][]{jahoda06} during the interval MJD
55482.008--55519.130 (ObsId.~95437). Spectra and response matrices
were extracted by using standard tools of the Heasoft software
package, version 6.11. The 'bright' model was used to subtract
background emission, and only photons recorded by the top layer of the
Proportional Counter Unit (PCU) 2 and falling in the 2.5 -- 30 keV
energy range were retained in spectral fitting.  Further, a systematic
error of 0.5\% was added to each spectral
channel\footnote{http://www.universe.nasa.gov/xrays/programs/rxte/pca/doc/rmf/pcarmf-11.7/}. {\it
  HEXTE} data were not included in the analysis since the background
dominates over source photons, and a number of 'line-like' residuals
of calibration origin make the characterisation of the spectral
continuum at high energies extremely problematic. In the temporal
analysis we retained photons collected by all the PCUs of PCA in
  the 2--60 keV energy band, and encoded in fast configurations such
as {\it Good Xenon} and {\it Event}, with temporal resolution of 1 and
122 $\umu$s, respectively.

\subsection{Swift}

The {\it Swift} observations of {\src} covered the interval MJD
55479.198--55501.559 (Obs.Ids 31838, 31841, 437313, 437466) and in
this paper we consider only data observed by the X-ray Telescope
\citep[XRT; 0.2--10 keV; ][]{gehrels04}. All the data were processed
by using standard procedures \citep{Bur05} and the latest calibration
files available at the time of the analysis (2011 September; caldb
v. 20110725).  Spectral analysis was performed on data collected both
in window timing (WT) and photon counting (PC) mode (processed with
the {\sc xrtpipeline} v.0.12.6), while only WT data sampled every 1.7
ms, and falling in the  0.2--10 keV, were considered for the
temporal analysis.  Filtering and screening criteria were applied by
using {\sc ftools} (Heasoft v.6.11).  We extracted source and
background light curves and spectra by selecting photons in the 1--10
keV energy band and event grades of 0 and 0-12, respectively for the
WT and PC mode.  Grades 1 and 2 in the WT data were filtered out in
order to limit the effect of the recently reported redistribution
problem on the source and background product (see the web-page
http://www.swift.ac.uk for details).  The only observation considered
that was performed in PC mode was affected by a strong pile-up, and
corrected according to the technique developed by \citet{vaughan06}.
Given the relatively high count-rate of the source during the WT
observations, we also checked if these data were affected by pile-up
by extracting for each of them a source spectrum from a circular
region in which we progressively excluded a larger and larger fraction
of the inner central pixels\footnote{see also
  http://www.swift.ac.uk/pileup.shtml}. Fits to these spectra did not
reveal any significant change in the model parameters, and thus we
assumed that XRT/WT data were not affected by pile-up. Spectral
channels were grouped to contain at least 50 counts each, as it was
done for all the spectra produced by the various instruments
considered in this paper, and a systematic error of 3\% was added to
each spectral bin.

\subsection{INTEGRAL}

\textsl{INTEGRAL} observations are commonly divided into ``science
windows'' (SCWs), i.e., pointings with typical durations of
$\sim$2-3~ks.  We considered all available SCWs that were performed in
the direction of the source during the outburst.  These comprised
publicly available observations of the Galactic bulge (GB) region in
satellite revolutions from 975 to 980 (ObsId. 0720001; MJD
55479.365--55491.940), and the data collected during the ToO
observation of PKS 1830--211 in rev. 981 (ObsId. 0770005; MJD
55494.778--55497.447).  During the GB observations, J17480 was
relatively close to the centre of the IBIS/ISGRI telescope
\citep[20--150 keV;][]{lebrun03,ubertini03} and thus it was
simultaneously observed also with the JEM-X telescope \citep[4.5--25
  keV;][]{lund03}.  During the observations performed in rev. 981, the
source was always outside the JEM-X field of view. All INTEGRAL data
were analysed using version 9.0 of the OSA software distributed by the
ISDC \citep{courvoisier03}.  JEM-X and ISGRI spectra were rebbinned in
order to have 16, 32 or 64 energy channels in the different
revolutions, depending on the available statistics of the data.

\section{Analysis}

\subsection{Spectral Analysis.}
\label{sec:spectrum}

\begin{figure}
  \includegraphics[angle=0,width=\columnwidth]{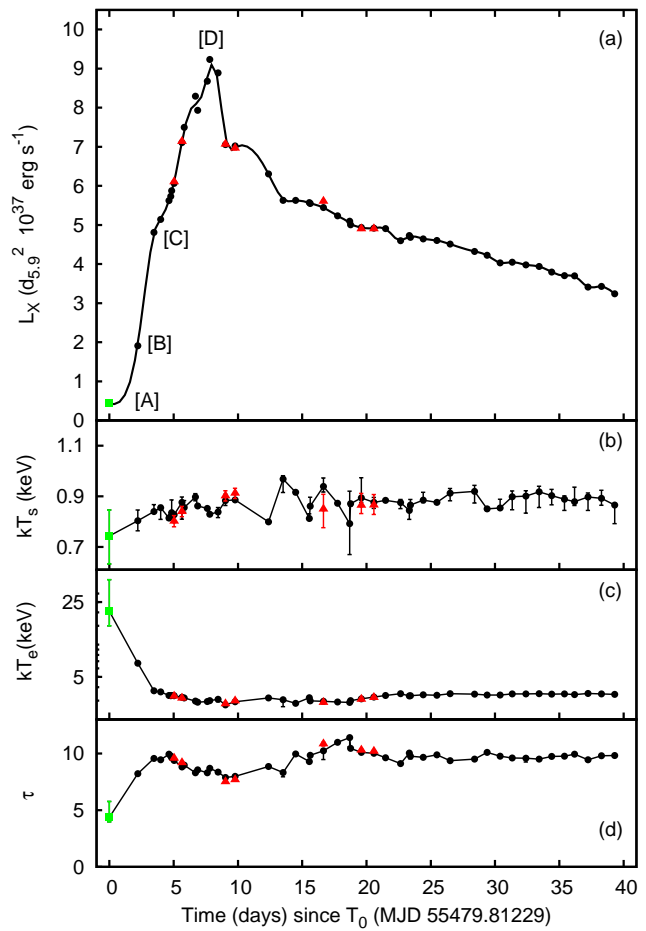}
 \caption{ (a) X-ray luminosity of {\src} estimated by extrapolating
   the best-fitting spectral model, \texttt{TBAbs*(nthcomp+gaussian)},
   to the 0.05--150 keV energy band.  Black points refer to the
   modelling of the observations performed by PCA on-board {\it RXTE},
   alone, red triangles to simultaneous XRT-{\it Swift} and PCA-{\it
     RXTE} observations, while green squares to simultaneous XRT-{\it
     Swift}, JEMX-\textsl{INTEGRAL} and ISGRI-\textsl{INTEGRAL} observations. The temperature of the
   soft input photons for Comptonization, the electron temperature of
   the Comptonizing region, and its optical depth, are plotted in the
   panels (b), (c) and (d), respectively.  Plotted errors are given at
   the 90\% confidence level. The solid line in panel (a) is a cubic
   spline approximation.  }
\label{fig:sp}
\end{figure}

The spectra of the 46 observations performed by PCA were modelled
using a thermal Comptonization model
\citep[\texttt{nthcomp,}][]{Zdz96,Zyc99}, with the addition of a
Gaussian emission line centred at energies compatible with Iron
K-$\alpha$ transition (6.4--6.97 keV). Interstellar absorption was
modelled using the \texttt{TBAbs} model \citep{Wlm00}, fixing the
absorption column to the value determined thanks to a 30 ks {\it
  Chandra} observation by \citet{Mll11},
$N_{\mbox{H}}=1.17\times10^{22}$ cm$^{-2}$. The assumed model
describes satisfactorily the observed spectra ($\chi^2_r\simeq1$; see
the rightmost column of Tab.~\ref{tab:xtespectra}, where the
parameters of the spectral continuum, fluxes in the 2.5--25 keV energy
band, and luminosity extrapolated to the 0.05--150 keV band, are also
reported for each spectrum).

\begin{figure}
  \includegraphics[angle=0,width=\columnwidth]{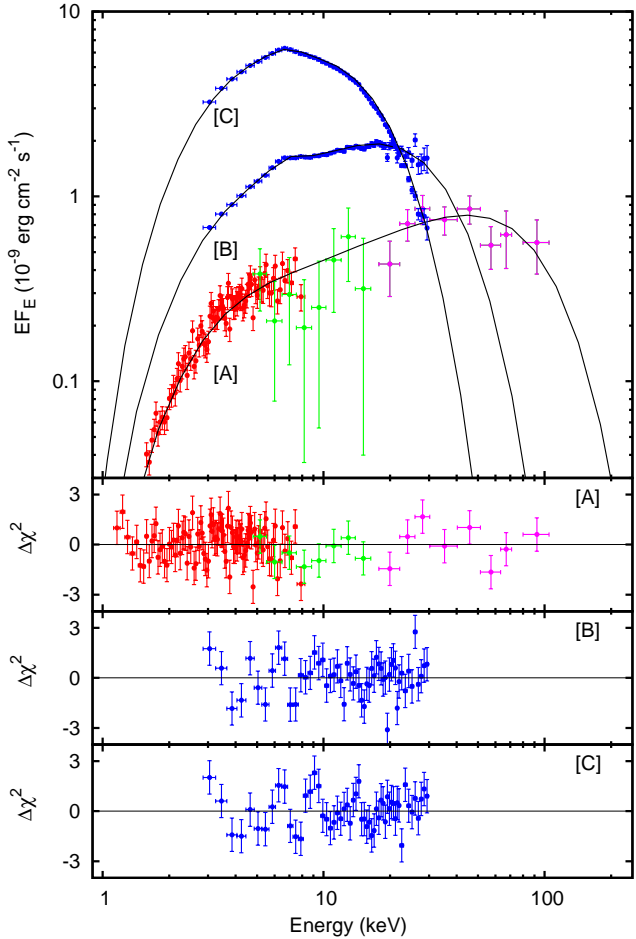}
\caption{ Unfolded spectra accumulated during \textsl{Swift} and
  \textsl{INTEGRAL} observations covering the intervals
  55479.801--55479.824 and 55479.3652–-55479.5483, respectively, [A],
  and during {\textsl{RXTE}} observations covering the intervals MJD
  55482.008--55482.043 [B] and MJD 55483.715--55483.839 [C], respectively. The
  best-fitting models, defined as \texttt{TBAbs*(nthcomp+gaussian)},
  are over-plotted as black solid lines. Points referring to XRT,
  JEM-X, ISGRI and PCA spectra are plotted as red, green, purple and
  blue points, respectively. In the lower panels, residuals with
  respect to the best-fitting models are plotted.
\label{fig:3sp}}
\end{figure}

 The \texttt{nthcomp} model describes a Comptonized spectrum in terms
 of the cloud electron temperature , $kT_e$, of the temperature of
 photons injected in the cloud, $kT_S$ and of an asymptotic power-law
 index, $\Gamma$, related to the cloud optical depth, $\tau$, by the
 relation
\begin{equation}
\label{eq:tau}
\Gamma=\left[ \frac{9}{4}+\frac{1}{\tau(1+\tau/3)(kT_e/m_ec^2)} \right]^{1/2}-\frac{1}{2},
\end{equation} \citep[see, e.g.,][]{LghZdz87}. 
The optical depth was then estimated by inverting such a relation,
  and the relative uncertainty estimated by propagating the
  statistical errors on $kT_e$ and $\Gamma$. The values of the
parameters of the Comptonized spectrum measured across the outburst ,
$kT_S$, $kT_e$, and $\tau$ are shown as black points in panels (b),
(c), and (d) of Fig.~\ref{fig:sp}, respectively. The 3--30 keV
spectrum of the source is relatively soft if compared with AMSP
\citep[see, e.g.,][and references therein]{Pou06}. With the exception
of the first \textsl{RXTE} observation (labelled as [B] in the top
panel of Fig.~\ref{fig:sp} and spanning the interval MJD
55482.008--5482.043), the electron temperature of the Comptonizing
medium is in fact relatively low ($\simlt 3.5$ keV) and the optical
depth relatively large ($\tau \approx$ 8--10).  By imposing energy
conservation between the cold ($kT_S \approx 0.9$ keV) and the hot
phase ($kT_e \approx 3$ keV during most of the observations) producing
the Comptonization spectrum, an estimate of the radius of the region
that provides the soft photons that are subsequently Comptonized in
the hot cloud is obtained. Since the total flux is $F=F_{S}(1+e^{y})$,
where $F_{S}=\sigma T_{S}^4 (R_{S}/d)^2$ is the observed flux coming
from the colder region and $y=4(kT_e/m_ec^2) \tau^2$ is the
Comptonization parameter, one has $R_{S}=1.8 (1/kT_{S})^2
\sqrt{F/(1+e^y)}$ d$_{5.9}$ km. Here, $kT_{S}$ is expressed in keV and
$d_{5.9}$ is the distance to the source in units of 5.9 kpc. By
inserting the observed values (see Tab.~\ref{tab:xtespectra}) an
emitting region with radius between 3 and 15 km is estimated,
indicating the NS surface as the most probable region of emission of
the soft photons. The iron line detected by PCA appears broad
($\sigma\approx$1 keV) and intense ($\mbox{EW}\approx0.2$--$0.3$ keV),
in accordance with the findings of \citet{Mll11} and
\citet{Ckr11}. Its energy is generally compatible with neutral iron
(6.4 keV), even if larger values are also sometimes observed.

\begin{figure}
  \includegraphics[angle=0,width=\columnwidth]{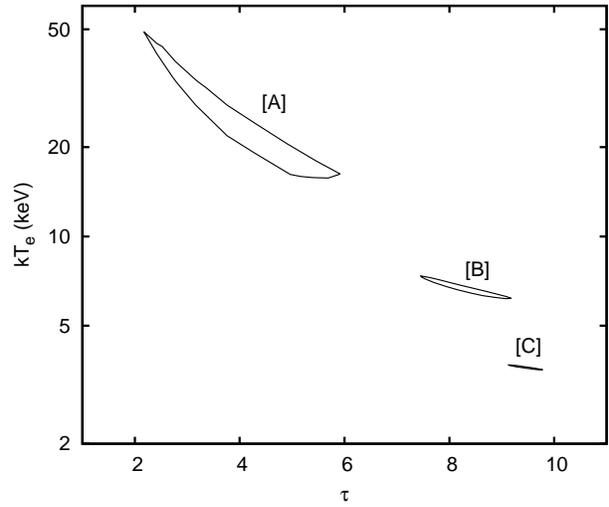}
\caption{Error contours on the electron temperature and optical depth
  measured during observations labeled as [A], [B] and [C] (see text
  and the caption of Fig.~\ref{fig:3sp}), drawn considering the 90\%
  confidence intervals calculated for two parameters
  ($\Delta\chi^2=4.61$).}
\label{fig:contours}
\end{figure}

The accuracy of the spectral modelling obtained from PCA observations
was checked by considering the spectra obtained by XRT during the six
{\it Swift} observations which partly overlapped or were very close in
time to {\it RXTE} pointings. During all of these observations XRT was
operated in WT mode.  Spectral fitting was performed by letting 
  the relative normalisation of XRT with respect to PCA free to vary;
  values in the range 0.92--0.96, with a typical uncertainty of 0.01,
  were measured.  The results obtained are given in
Tab.~\ref{tab:swxtespectra}. The combined spectra are well described
by the model discussed previously, even if residuals are sometimes
found at the overlap between the two energy bands.  The best-fitting
parameters of the simultaneous modelling of XRT and PCA spectra are
plotted as red points in Fig.~\ref{fig:sp}, and are visibly very close
to the values obtained by modelling PCA spectra alone.  The addition
of XRT spectra then confirms that the spectral decomposition obtained
by analysing PCA data alone is robust.  The values of the absorption
column obtained with the simultaneous modelling of XRT and PCA spectra
are between 1.0 and 1.5$\times10^{22}$ cm$^{-2}$, compatible with the
{\it Chandra} estimate.

As it was noted earlier, the source emission is soft during the large
part of the outburst. Nevertheless, the electron temperature measured
during the first PCA observation (labelled as [B] in the top panel
Fig.~\ref{fig:sp}), 6.7(2) keV, suggests that a spectral transition
from a harder state took place during the outburst rise.  To show this
spectral change we plot in the top panel of Fig.~\ref{fig:3sp} the
unfolded spectra extracted during observations designated as [B]
(covering the interval MJD 55482.008--5482.043) and [C] (MJD
55483.715--55483.839; see top panel of Fig.~\ref{fig:sp}). To further
investigate this spectral transition, we also considered observations
performed by {\it Swift} and {\it INTEGRAL} before the start of the
{\it RXTE} coverage on MJD 55482.008 (see label [A] in the top panel
of Fig.~\ref{fig:sp}).  We then fitted the 1--10 keV {\it Swift} XRT
spectrum accumulated starting on MJD 55479.801, for an exposure of 2.0
ks (Obs. 00031841002), together with the 4.5--20 keV 
  JEMX-\textsl{INTEGRAL} and the 20--150 keV
  ISGRI-\textsl{INTEGRAL} spectra accumulated during Rev.975
(spanning MJD 55479.365 -- 55479.548, for an exposure of $\sim 8.5$
ks). We normalised the flux observed by JEM-X and ISGRI to that
  observed by XRT by letting a normalisation constant free to
  vary. Given the relatively low count rates observed by JEM-X and
  ISGRI, we held the normalisation constant between these two
  instruments fixed to 1. No significant improvement in the modelling
  is obtained otherwise. The spectrum so produced is well fitted by
the same Comptonization model used above, with best-fitting
parameters, $kT_{S}=0.7(1)$ keV, $kT_e=20_{-5}^{+20}$ keV and
$\tau=4.4_{-0.4}^{+1.4}$ (see Table \ref{tab:obsAspectra}). These
indicate a hotter and optically thinner Comptonizing medium than what
is observed during later observations. This is clearly shown in the
top panel of Fig.~\ref{fig:3sp}, where the unfolded spectrum of the
combined XRT, JEM-X and ISGRI spectra is labelled as [A].  The
best-fitting parameters of spectrum [A], as well as the luminosity
extrapolated to the 0.05-150 keV energy band, are reported in
Fig.~\ref{fig:sp} with green points. To better show the spectral
  transition performed by {\src} during the outburst rise, and to take
  into account the correlation between the values of electron
  temperature and optical depth measured with the Comptonization model
  we employed (see Eq.~\ref{eq:tau}), we plot in
  Fig.~\ref{fig:contours} the 90\% confidence level error contours for
  these parameters. The contours were obtained by considering the
  surface for which the fit chi-squared was larger by
  $\Delta\chi^2=4.61$ with respect to the best fitting value.

\begin{figure}
  \includegraphics[angle=0,width=\columnwidth]{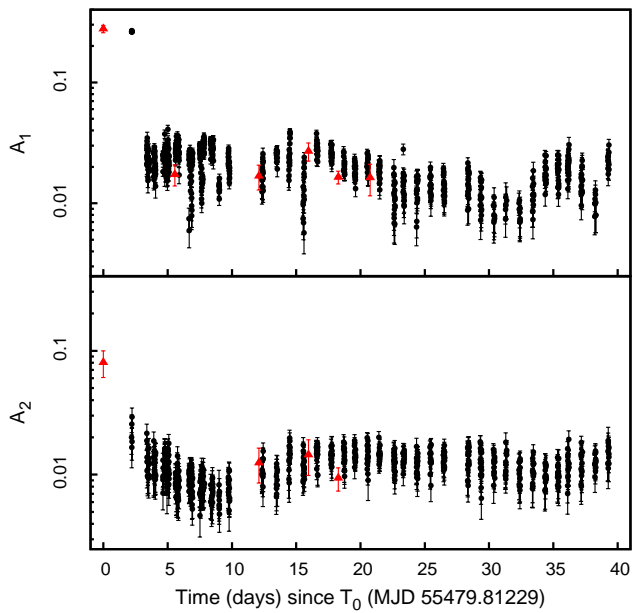}
 \caption{Background-subtracted fractional amplitude of the first
   (top) and second (bottom) harmonic component used to model the
   pulse profile of {\src}. Red triangles refer to profiles obtained
   by folding 0.2--10 keV XRT time series around
   $\nu_0=11.044885$ Hz over the entire length of each observation,
   while black points are obtained by folding 300 s long segments of
   2--60 keV PCA data around the same frequency. }
\label{fig:ampl}
\end{figure}

A softening of the emission during the outburst rise is also indicated
by the analysis of the spectra accumulated by JEM-X and by ISGRI
on-board \textsl{INTEGRAL}, alone. We list in
Tab.~\ref{tab:igrspectra} the results obtained by fitting the spectra
accumulated during each of the \textsl{INTEGRAL} revolution with the
same model used so far, and keeping the relative normalisation
  between the two instruments fixed to one. Observations performed
before MJD 55483 are well fitted by Comptonization in a medium with
$kT_e>10$ keV, while subsequent observations indicate temperatures of
the order of those indicated by the analysis of XRT and PCA spectra
($\approx3$ keV). Such a transition is even clearer by taking into
account the fluxes observed by JEM-X and ISGRI. Before MJD 55483 the
ratio $F_{JEMX}(4.5-25.0)/F_{ISGRI}(20.0-150)$ is $\approx 0.5-1$,
while it takes values in excess of 30 during observations performed
when the source bolometric luminosity is larger. Moreover ISGRI could
not detect the source after MJD 55483, with typical upper limits of
$F_{ISGRI}(50-150)<\mbox{few}\:\times\:10^{-11}$--$10^{-12}$ erg
cm$^{-2}$ s$^{-1}$ (3 $\sigma$ confidence level), while the flux in
the same energy band is $\approx 5\times10^{-10}$ erg cm$^{-2}$
s$^{-1}$ during earlier observations (see the rightmost columns of
Tab.~\ref{tab:igrspectra}).

\begin{figure}
 \includegraphics[angle=0,width=\columnwidth]{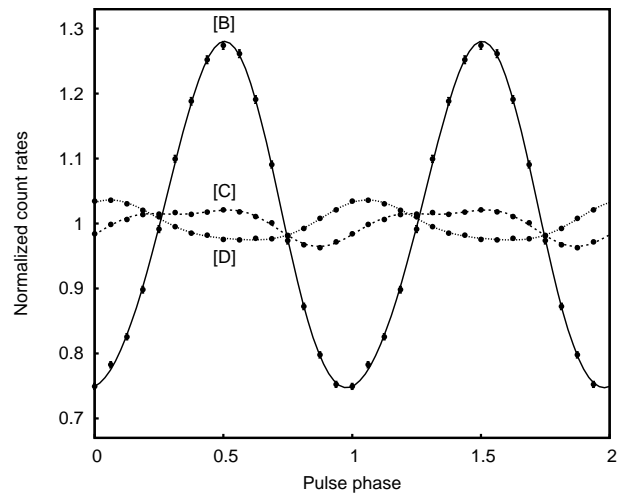}
 \caption{ Background-subtracted profiles obtained by folding around
   $\nu_0=11.044885$ the 2--60 keV PCA light curves accumulated
   during observation labelled as [B] (spanning MJD
   55482.008--55482.043), [C] (MJD 55483.715--55483.839) and [D] (MJD
   55487.565 -- 55487.693), when the 0.05--150 keV luminosity was
   $1.91(4)\times10^{37}$, $4.81(2)\times10^{37}$ and
   $9.23(4)\times10^{37}$ erg s$^{-1}$, respectively. The difference
   of the phase of the various pulse profiles is interpreted in
   Sec.~\ref{sec:timing} and \ref{sec:torque} in terms of the NS
   spin frequency derivative. Two cycles are plotted for clarity.}
   \label{fig:sequenzaprofili}
\end{figure}

Estimates of the bolometric luminosity were obtained by extrapolating
the best-fitting models of PCA spectra to the 0.05-150 keV energy
band, and using the most updated estimate of the distance to the
cluster (d=5.9$\pm$0.5 kpc, \citealt{Lnz10}). The values we obtained
are listed in Tab.~\ref{tab:xtespectra} and plotted in panel (a) of
Fig.~\ref{fig:sp}. The maximum luminosity, $L_{max}=9.23(4)\times10^{37}$
d$_{5.9}^2$ erg s$^{-1}$, is observed during the observation starting
on $\bar{t}=\mbox{MJD}\:55487.63(6) = 7.81(6) $ d since $T_0$, and
labelled as [D] in the top panel of Fig.~\ref{fig:sp}.  To give an
estimate of the X-ray flux before the start of the {\it RXTE} coverage
of the outburst, we consider XRT and \textsl{INTEGRAL} observations
previously labelled as [A], yielding an estimate
$L_{min}=0.44(6)\times10^{37}$ d$_{5.9}^2$ erg s$^{-1}$.

\subsection{The pulse profile}
\label{sec:pulse}

To analyse the properties of the pulse profile of {\src} we considered
the time-series recorded by PCA in the 2--60 keV energy band and
by the XRT in the 0.2--10 keV band. We first applied barycentric
corrections on the times of arrival of photons recorded by PCA and
XRT, to take into account the motion of the spacecrafts in the Solar
System. To this end we considered the position, RA=17$^{h}$ 48$^{m}$
4$\fs$831(4), DEC=-24$^{\circ}$ 46' 48$\farcs$87(6), determined by
\citet[][]{Poo10} by comparing a {\it Chandra} observation of the
source during the outburst with a deeper past {\it Chandra}
observation of Terzan 5, performed when {\src} was quiescent
\citep{heinke06}. This postion is compatible with the one derived
  by \citet{riggio2012} from a Moon occultation of the transient,
  observed by \textsl{RXTE} during the first days of the outburst.
The NS motion in the binary system was corrected by using the orbital
parameters found by P11, who performed a timing analysis on a subset
of the observations considered here.

PCA time series were split in 300 s long intervals and folded in 16
phase bins around $\nu_0=11.044885$ Hz (see P11).  We adopt the
criterion given by \citet{Leh87} to determine a detection threshold.
A signal was detected with a significance larger than 3 $\sigma$
in all PCA observations. Given its lower effective area, the
count-rates recorded by XRT are significantly lower and the whole
length of each observation was considered in the folding procedure to
maximise the counting statistics. A signal could be detected only in 6
out of 18 XRT observations; however, when a detection
was missing the upper limit on the signal amplitude was larger than
the amplitude indicated by the analysis of quasi-simultaneous PCA
observations. The limited signal-to-noise ratio of XRT
data sets is then the most plausible reason for these non-detections.

\begin{figure}
 \includegraphics[angle=0,width=\columnwidth]{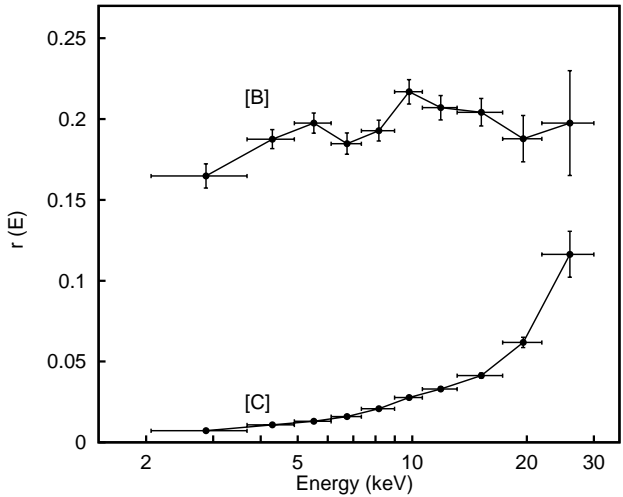}
 \caption{Energy dependence of the RMS amplitude of the pulse profile
   shown by {\src} during PCA observations performed during the
   interval MJD 55482.008-55482.043 (labelled as [B]) and MJD
   55483.715--55483.839 (labelled as [C]), respectively.}
   \label{fig:rms_ampl}
\end{figure}

A two-components harmonic decomposition well describes the observed
pulse profiles.  The evolution of the background-subtracted 2--60 keV
fractional amplitude of the first ($A_1$) and second harmonic ($A_2$)
observed by PCA is plotted as black points in the top and bottom panel
of Fig.~{\ref{fig:ampl}}, respectively.  As it was already noted by
\citet{Pap11}, a sudden variation in the pulse profile properties
takes place during the outburst rise. Most noticeably, the fractional
amplitude of the first harmonic observed by PCA decreases from values
of $\sim 27\%$, measured before MJD 55483, to $\sim 2\%$ during later
observations.  The second harmonic amplitude is also observed to
decrease, though much more smoothly than the amplitude of the
fundamental. During the decaying stage of the outburst, the amplitudes
of the two harmonic components do not show any other step-like
behaviour, following a smooth trend. Even though the different
  instrumental responses do not allow an immediate comparison, also
the amplitude observed by XRT in the 0.2--10 keV energy band follows a
similar trend (see red triangles in Fig.~{\ref{fig:ampl}).

  To show the evolution of the pulse profile shape taking place around
  MJD 55483 we plot in Fig.~\ref{fig:sequenzaprofili} the
  background-subtracted profiles observed by PCA during the
  observation labelled as [B], [C] and [D], covering the intervals MJD
  55482.008-55482.043, MJD 55483.715--55483.839 and MJD
  55487.565--55487.693, and during which the 0.05-150 keV X-ray
  luminosity was $1.91(4)\times10^{37}$, $4.81(2)\times10^{37}$ and
  $9.23(4)\times10^{37}$ erg s$^{-1}$, respectively.

\begin{figure}
 \includegraphics[angle=0,width=\columnwidth]{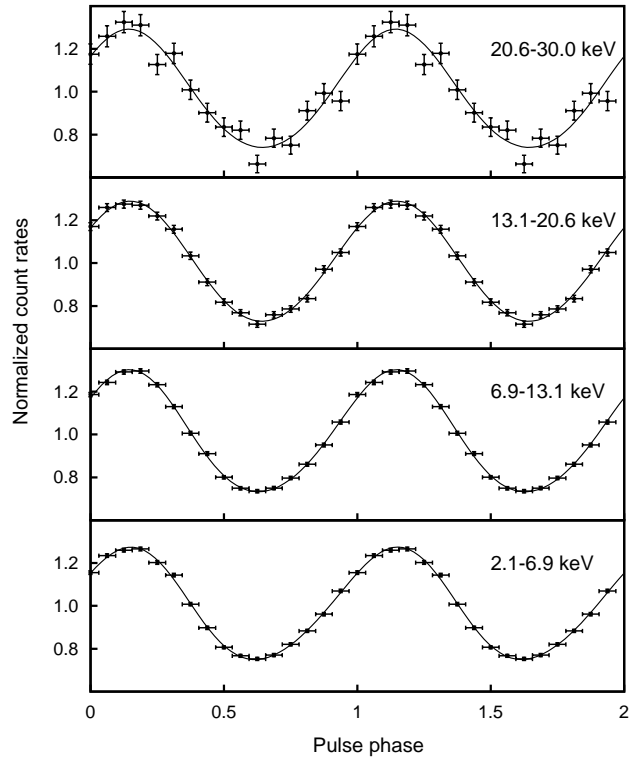}
 \caption{Background subtracted pulse profiles in four energy bands
   accumulated during PCA observations of group [B] (see
   text). The solid curves are the respective best-fitting harmonic
   decompositions. Two cycles are plotted for clarity.}
   \label{fig:4profiles_55482}
\end{figure}

Also the energy dependence of the pulse amplitude changes during the
outburst rise. To show this, we consider the observations performed by
{\it RXTE} during the intervals [B] and [C]. Such a choice is made
since the pulse properties observed during group [C] data
qualitatively represent what is observed during the rest of the
outburst.  The energy dependence of the (background subtracted) pulse
RMS amplitude, $r(E)=[(A_1^2+A_2^2)/2]^{1/2}$, is plotted in
Fig.~\ref{fig:rms_ampl}. During observations labelled as [B] the pulse
RMS amplitude is roughly constant with an average value of
$0.19$. After MJD 55483 instead, the pulses are nearly completely
suppressed at low energies and become stronger as the energy increases
up to 30 keV (see curve labelled as [C] in Fig.~\ref{fig:rms_ampl}).
While during observations of group [B] the profile is nearly
sinusoidal and shows the same shape in different energy bands, the
phase of the two harmonics becomes slightly energy dependent after MJD
55483. To see this, we plot in Fig.~\ref{fig:4profiles_55482} and
\ref{fig:4profiles} the background subtracted pulse profiles in four
PCA energy bands, accumulated during observations labelled as [B] and
[C], respectively. During observations of group [B] the profile has a
small harmonic content and shows a similar shape in different energy
bands. During observations [C] the phase difference between the first
and the second harmonic is negligible at low energies, resulting in a
nearly symmetric profile shape, whereas, at higher energies, the
fundamental lags the second harmonic resulting in a more skewed
profile.

\begin{figure}
 \includegraphics[angle=0,width=\columnwidth]{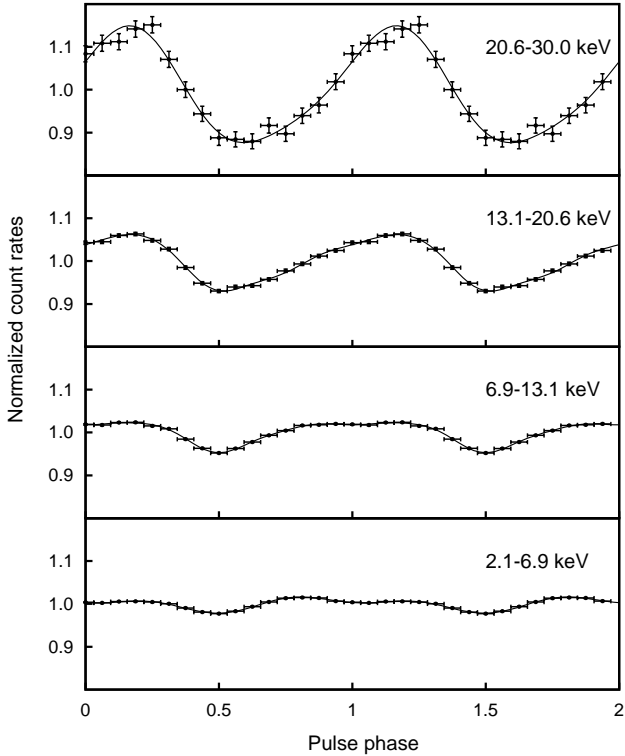}
 \caption{Background subtracted pulse profiles in four energy bands
   accumulated during PCA observations of group [C] (see
   text). The solid curves are the respective best-fitting harmonic
   decompositions. Two cycles are plotted for clarity.}
   \label{fig:4profiles}
\end{figure}

\subsection{Timing analysis.}
\label{sec:timing}
The spin evolution of the source during the outburst was investigated
under the assumption that the evolution of the phase of the pulse
profile is a good proxy of the NS spin, that is
\begin{equation}
\phi(t)-\phi_0=-\int_{T_0}^t [\nu(t')-\nu_f] dt',
\label{eq:phase1}
\end{equation}
where $\nu(t)$ is the spin frequency of the NS, $\nu_f$ is the
frequency around which the time series are folded and $T_0$ = MJD
55479.81229 is the reference epoch of the timing solution. The phases
calculated over the first and the second harmonic component are
plotted in the top panels of Fig. \ref{fig:phase1} and
\ref{fig:phase2}, respectively. Phases computed over the profiles
observed by XRT in the 0.2--10 keV do not show a signficant scatter
with respect to PCA phases (2--60 keV). This can be seen from the
bottom panels of Fig. \ref{fig:phase1} and \ref{fig:phase2}, where we
plotted the residuals of XRT (red triangles) and PCA (black points)
phases with respect to the best-fitting models defined below.  We
conclude that the influence on the observed phases of the different
energy band and response of the two instruments is small, and consider
both the values measured by XRT and PCA. Considering PCA phases alone
would not significantly alter the results obtained.

\begin{figure}
 \includegraphics[angle=0,width=\columnwidth]{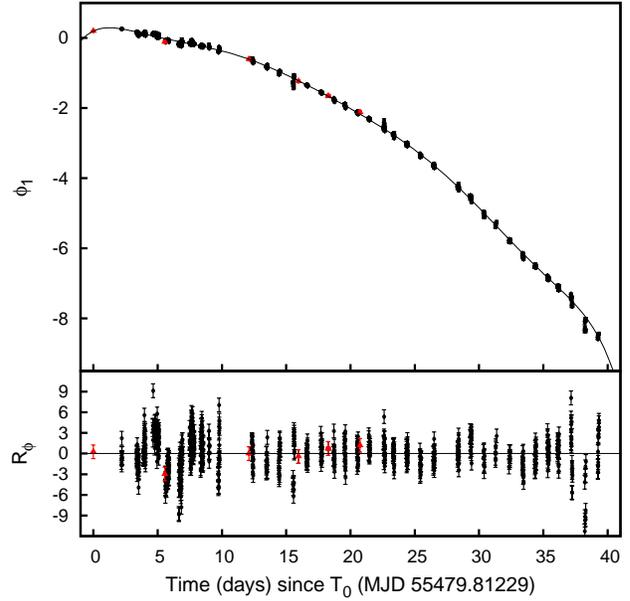}
 \caption{Phases computed over the first harmonic component by folding
   PCA (2--60 keV; black points) and XRT (0.2--10 keV; red
   triangles) time series around $\nu_0=11.044885$ Hz, together with
   the best fitting 8-th order polynomial (see text) plotted as a
   solid line (top panel). Residuals in units of sigma with respect to
   the best fitting model are plotted in the bottom panel. }
\label{fig:phase1}
\end{figure}

To characterise phenomenologically
their evolution we fit them with the relation
\begin{equation}
\phi(t)-\phi_0=P_N(t-T_0)+R_{orb}(t-T_0).
\label{eq:1}
\end{equation}
Here, $P_N(t-T_0)=\sum_{n=0}^{N} c_n p_n(t-T_0)$ is a $N$-th order
polynomial expansion, and the $p_n$'s are discrete orthogonal
polynomials over the considered data set. Such a choice makes
the various fitted parameters, $c_n$, independent on the degree of the
fitted polynomial, $N$ \citep[see, e.g.,][]{Bev}. Defining
$x_i=t_i-T_0$, the terms of the polynomial expansion are obtained
through the recurrence relation,
$p_{n+1}(x)\equiv(x-A_{n+1})p_n(x)-B_np_{n-1}(x)$, with $p_0(x)=1$ and
$p_{-1}(x)=0$. The set of constants $(A_n,B_n)$ is determined by
imposing orthogonality between the various terms of the polynomial
expansion, i.e. $\sum_i (1/\sigma_i^2)p_k(x_i)p_j(x_i)=0$, for $k\neq
j$, $k,j=0,...,N$. The weights $1/\sigma_i^2$ simply reflect the
uncertainties on the various phase points. The term $R_{orb}(t)$ is
added to Eq. \ref{eq:1} to take into account the effect on the phases
of a small difference between the orbital parameters used to correct
the time series (the orbital period $P_{orb}$, the projected
semi-major axis $a\,\sin{i}$ and the epoch of zero mean longitude
$T^*$) and the actual orbital parameters of the system. An expression
for $R_{orb}(t)$ is given by \citet{deeter81}.

\begin{figure}
 \includegraphics[angle=0,width=\columnwidth]{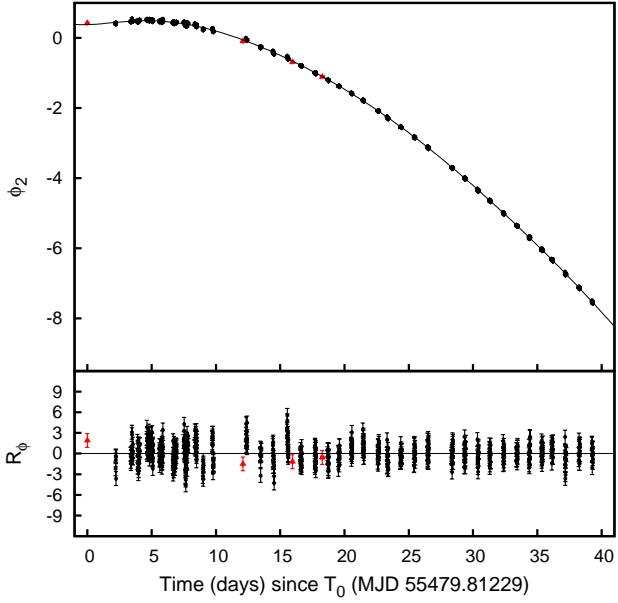}
 \caption{Same as Fig. \ref{fig:phase1} regarding the phases computed
   over the second harmonic component.}
   \label{fig:phase2}
\end{figure}

Small corrections to the orbital parameters are found, leading to the
estimates given in Table. \ref{table:orbit} (obtained considering a
fit with $N=8$, see below).  The relatively large values the
$\chi^2_r$ obtained from the phase modelling indicates the presence of
phase noise.  Slight differences are found between the orbital
parameters obtained by fitting the two different harmonics, indicating
that such a noise induces phase variations also on a timescale
comparable with the orbital period of the system.  To give
conservative estimates of the orbital parameters we then quote the
average among the parameters obtained from the phases of the first and
second harmonic. The uncertainty on each parameter is estimated to
encompass the 1 $\sigma$ confidence level interval of the value
indicated by the analysis of each component.  The values of the
orbital parameters we obtain are compatible with the estimates
obtained by P11 on a reduced data set. The new set of 
parameters  is then used to correct the time series. The
errors induced on the phase estimates by the uncertainties in the
orbital parameters [see Eq. 3 in \citet{Brd07}] are now summed in
quadrature to the statistical phase uncertainties. By doing so any
residual orbital modulation of the pulse phases due to the uncertainty
affecting the orbital parameters is already included in the phase
errors. The fitting of the phases is then repeated by using
Eq. \ref{eq:1} with $R_{orb}(t)=0$. This makes the function
defined by Eq.~\ref{eq:1} eventually orthogonal on the considered data
set.

Starting from $N=1$, the order of $P_N(t-T_0)$ is increased and the
phases re-fitted with the higher-order polynomial until the
improvement in the $\chi^2$ of the fit is not significant on a 3 $\sigma$
confidence level according to an F-test, after two consecutive
order-increments. Despite that such a criterion is somewhat arbitrary,
the orthogonality between the various orders of the polynomial
expansion grants that the estimates of the fitted parameters, $c_j$,
$j\le N$, do not depend on the polynomial order, allowing a coherent
phenomenological description of the phase evolution.

A by-eye inspection of Fig. \ref{fig:phase1} and Fig. \ref{fig:phase2}
already shows the need for a quadratic term to fit the phase
evolution. This is further confirmed by the extremely large values of
the F function ($>10^5$) obtained when adding such a term to a simple
linear function.  Under the assumption that the phases are a good
tracker of the NS spin (Eq. \ref{eq:phase1}), we have
\begin{equation}
\label{eq:derivatives}
\frac{d^{(n)}}{dt^{(n)}}[\phi(t)-\phi_0]=-\frac{d^{(n-1)}}{dt^{(n-1)}}\nu(t),
\end{equation} 
which shows how the best-fitting value of every parameter $c_n$
translates into an estimate of the average value of the (n-1)-th
derivative of the spin frequency. The values of the spin frequency at
$T_0$ and of the average spin frequency derivative we obtain from the
fit of the first and the second harmonic phases are quoted in
Table~\ref{table:freq}. The two estimates of the average spin
frequency derivative are very close to each other, and a value of
$<\dot{\nu}>=1.48(2)\times10^{-12}$ Hz s$^{-1}$ encompasses both.  By
considering a smaller dataset (the first 25 days of the outburst
compared to the 40 days considered here), P11 obtained values of the
average spin-up rate of 1.22(1) and 1.68(1)$\times10^{-12}$ Hz
s$^{-1}$, from the phases computed over the first and the second
harmonic, respectively. This clearly shows how the considered dataset
may influence the determination of the spin parameters of an accreting
pulsar when profile it shows is affected by timing noise. However, the
scatter between the values obtained from the first and second harmonic
greatly decreases when a longer interval is considered, suggesting how
the estimate given here is more reliable.

\begin{figure}
 \includegraphics[angle=0,width=\columnwidth]{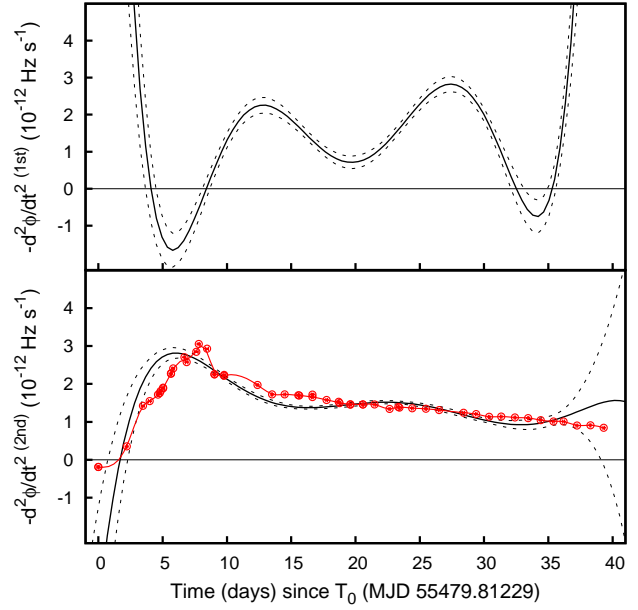}
 \caption{Black solid lines represent the second derivative of the
   8-th order polynomial that best-fits the phases computed over the
   first (top panel) and the second harmonic (bottom panel). Dashed
   black curves mark the respective 1$\sigma$ confidence-level
   intervals. The change of sign of the second derivative of the
   best-fit polynomial is done to make the plotted function equal to
   the spin frequency derivative (see Eq.~\ref{eq:derivatives}) and to
   facilitate comparison with light curve plotted in the top panel of
   Fig.~\ref{fig:sp} and reproduced in the bottom panel of
     this figure by using red circles (flux units are arbitrary).}
   \label{fig:deriv}
\end{figure}

The phase evolution is anyway far from being adequately fitted by a
simple quadratic function and significant improvements of the
description are successively obtained adding higher order terms, up to
$N=8$. Considering that the phases computed over the two harmonic
components have similar uncertainties, the quality of the modelling of
the second harmonic phases ($\chi^2_r=1371.8/673=2.04$) is
significantly better than for the first one
($\chi^2_r=4614.2/760=6.07$). Yet, these values indicate the presence
of phase fluctuations which are not described by a polynomial
expansion of a relatively low order. These fluctuations affect in a
different manner the two harmonic components, though, and the trend
followed by second harmonic phases is smoother than that followed by
the phases computed over the first harmonic.  To quantify this we
evaluated the total RMS amplitude of the phase residuals with respect
to the best-fitting polynomial as the squared sum of the statistical
uncertainty on the phase measurement (which depends on the amplitude
of the considered harmonic and on the count rate) and of the intrinsic
phase noise, $\sigma^2_{k,rms}=\sigma^2_{k,stat}+\sigma^2_{k,noise}$,
where $k$ is the harmonic number \citep[see][and references
  therein]{H08}. By taking an average value for the statistical
uncertainties one obtains that $\sigma_{1,noise}=0.055$ and
$\sigma_{2,noise}=0.017$.

\begin{table}
\caption{Orbital parameters of {\src} obtained by modelling with
  Eq. \ref{eq:1} the phases computed on the first (left column) and on
  the second (central column) harmonic, together with the $\chi^2$ of
  the respective modelling. Uncertainties are given at 1$\sigma$
  confidence level and were scaled by a factor $\sqrt{\chi_r^2}$ to
  take into account that $\chi^2>1$. The average between the values
  obtained considering the two harmonic components are listed in the
  rightmost column, with the uncertainties estimated conservatively
  such as to overlap the values indicated by the two components. }
\label{table:orbit}
\centering
\renewcommand{\footnoterule}{} 
\begin{tabular}{lrrr}
\hline
         & 1$^{st}$ harmonic & 2$^{nd}$ harmonic & Average\\
a $\sin{i}$/c  (lt-s) & 2.4972(2) & 2.4972(1) & 2.4972(2) \\
P$_{orb}$ (s) & 76588.4(1) & 76587.92(6) & 76588.1(3) \\
T$^*$ (MJD) & 55481.78038(2) & 55481.78052(1)&	55481.78045(8)\\
e & $<6\times10^{-4}$& $<1\times10^{-4}$ & $<2\times10^{-4}$ \\
\hline
$\chi^2$ & 5192.99/757&1436.74/668 & \\
\hline
\end{tabular}
\end{table}

An enhanced stability of the second harmonic phases is also indicated
by the smooth trend followed by the second derivative of the best-fit
polynomial, which equals in modulus the spin frequency derivative
under the assumption that the NS spin is well tracked by the pulse
phases (see Eq.~\ref{eq:derivatives}). In the top and bottom panel of
Fig.~\ref{fig:deriv} we use black solid lines to plot the second
derivative of the best-fitting function of the phases computed over
the first and second harmonic, respectively.  The curve computed over
the first harmonic phases shows an oscillatory behaviour, especially
towards the boundaries of the available data set.  Such rapid
variations of the spin up rate in absence of simultaneous swings of
the X-ray light curve (the shape of which is reproduced in the bottom
panel by using red circles; see also the top panel of
Fig.~\ref{fig:sp}) indicate how hardly the phase variations computed
over the fundamental frequency component can be interpreted in terms
of a variable torque acting onto the NS (see
Sec.~\ref{sec:disc_torque}).  This possibly indicates that the
amplitude of the timing noise affecting the phases evaluated over this
component is larger than the deviations from a parabolic trend due to
the non constancy of the accretion torque, making the latter
practically unmeasurable from this harmonic component.  The phases
computed over the second harmonic have instead a much smoother
behaviour. The spin up rate increases during the first days reaching a
maximum value of 2.8(1)$\times$10$^{-12}$ Hz s$^{-1}$ at the epoch
6.0$\pm$0.5 d since $T_0$, and subsequently smoothly decreases until
it becomes ill-defined towards the boundary of the given
data-set. Comparing the spin up rate calculated over the second
harmonic phases to the X-ray light curve, it appears clear that the
$\dot{\nu}$(t) curve follows much more closely the expected dependence
on the instantaneous mass accretion rate.

These results indicate how the phases computed over the second
harmonic are possiby a better tracker of the spin frequency of
the NS.  For this reason we use only this component to investigate if
the observed behaviour is compatible with the accretion of Keplerian
angular momentum of the matter at the inner edge of the disc (see next
section).  This choice is further motivated by the analogous behaviour
observed in other accreting pulsars, where the second harmonic phases
are often observed to be less affected by timing noise than the
fundamental phases (see the Discussion, where a plausible explanation
for such a behaviour is also proposed).

\subsection{Torque modelling}
\label{sec:torque}

To analyse whether the evolution of the pulse phases computed over the
second harmonic can be described in terms of the torque imparted to
the NS by the accreting matter, we assume that the torque depends on a
power, $\beta$, of the mass accretion rate, $2\pi I_* \dot{\nu}\propto
\dot{M}^{\beta}$, and that the mass accretion rate is well tracked by
the X-ray luminosity, $\dot{M}=R_* L_x /\eta G M_*$, where $M_*$,
$R_*$ and $I_*$ are the mass, radius and moment of inertia of the NS,
respectively, and $\eta\simeq1$ is the accretion efficiency. We thus
consider a relation
\begin{equation}
\dot{\nu}(t)=\dot{\nu}(\bar{t})\left[\frac{L_X(t)}{L_X(\bar{t})}\right]^{\beta}
\label{eq:torque}
\end{equation} 
to express the dependence of the spin frequency derivative on the
accretion luminosity.  The bolometric accretion luminosity is
estimated by extrapolating to the 0.05--150 keV energy band the
best-fitting spectral models discussed in Sec.~\ref{sec:spectrum}. The
epoch $\bar{t}=\mbox{MJD}\:55487.63(6)=7.81(6)$ d since $T_0$, is the
epoch at which the maximum luminosity,
$L_X(\bar{t})=L_{max}=9.23(4)\times 10^{37}$ d$_{5.9}^2$ erg
s$^{-1}$ is attained.

\begin{table}
\caption{Frequency at the reference epoch and average spin frequency
  derivative obtained by fitting the first and second harmonic phases
  with the 8-th order polynomial defined by Eq.~\ref{eq:1} (with
  $R_{orb}(t-T_0)=0$), together with the relative $\chi^2$ and amplitudes
  of residuals (see text). Uncertainties are given at
  1$\sigma$ confidence level and were scaled by $\sqrt{\chi_r^2}$ to
  take into account that $\chi^2_r>1$.}
\label{table:freq}
\centering
\renewcommand{\footnoterule}{} 
\begin{tabular}{lrr}
\hline
         & 1$^{st}$ harmonic & 2$^{nd}$ harmonic\\
$\nu(T_0)$ (Hz) &11.0448803(7) & 11.0448797(5) \\
$<\dot{\nu}>$ (Hz s$^{-1}$) &1.489(4)$\times10^{-12}$ & 1.468(2)$\times10^{-12}$ \\

\hline
$\sigma_{rms}$ & 0.059& 0.024 \\
$\sigma_{stat}$ & 0.022& 0.017\\
$\sigma_{noise}$ & 0.055& 0.017\\
\hline
$\chi^2$ &4614.2/760 &1371.8/673  \\
\hline

\end{tabular}
\end{table}

If the folding
frequency is close to the actual spin frequency of the source and the
frequency variation throughout the outburst is small,
Eq.~\ref{eq:phase1} can be expressed as
\begin{eqnarray}
&\phi(t)-\phi_0 \simeq -\delta\nu\:(t-T_0)-\int_{T_0}^t dt'
  \int_{T_0}^{t'} \dot{\nu}(t'')\:dt''={}\nonumber\\ &{}=
  -\delta\nu\:(t-T_0)-\dot{\nu}(\bar{t})\:\int_{T_0}^t dt'
  \int_{T_0}^{t'}[L_X(t'')/L_X(\bar{t})]^{\beta}\:dt'',
\label{eq:fit}
\end{eqnarray}
where $\delta\nu=\nu(T_0)-\nu_f$, and we have inserted the assumed
dependency of the frequency derivative on X-ray luminosity
(Eq.~\ref{eq:torque}). One thus obtains a relation that can be used to
fit the observed phase evolution depending on the parameters $\phi_0$,
$\nu(T_0)$, $\dot{\nu}({\bar{t}})$ and $\beta$.

In practise, we approximated the luminosity profile of the outburst
$L_X(t)$ by using a natural smoothing cubic spline, which is plotted
as a solid curve in panel (a) of Fig.~\ref{fig:sp}. We then performed
numerically the double integral of the function $L_X^{\beta}(t)$ for a
set of values of $\beta$. We thus obtained a set of functions that
were used to fit the observed evolution of the phases computed over
the second harmonic. The best-fit is found for $\beta=1.07\pm0.03$
with a $\chi^2=1574/672=2.34$, a value not far from the $\chi^2$
of the best-fitting 8-th order polynomial ($\chi^2_r=2.04$). This
suggests a link between the phase variation computed over the second
harmonic and the X-ray flux measured from the source.  The
best-fit values of the spin frequency at the beginning of the
outburst $T_0$ and the spin frequency derivative at the outburst peak
are, $\nu(T_0)=11.044884819(6)$ Hz and
$\dot{\nu}(\bar{t})=2.755(5)\times10^{-12}$ Hz s$^{-1}$,
respectively. The errors on the fitted parameters (quoted at the
1$\sigma$ confidence level) were evaluated by scaling the statistical
errors by $\sqrt{\chi^2_r}$ to take into account the fact that
$\chi^2_r>1$. The best-fitting model is plotted over measured data
points, together with residuals, in Fig.~\ref{fig:beta11}.

\begin{figure}
 \includegraphics[angle=0,width=\columnwidth]{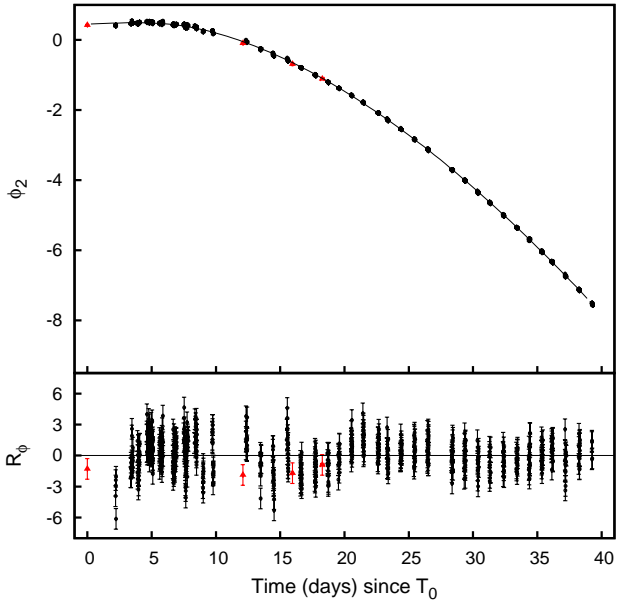}
 \caption{Best fitting model according to the physically-motivated
   torque model defined by Eq.~\ref{eq:torque} and \ref{eq:fit} of the
   second harmonic phases (top panel), together with residuals in units of
   $\sigma$ (bottom panel). }

   \label{fig:beta11}
\end{figure}

To interpret the timing noise affecting the pulse phases observed from
AMSP, \citet{patruno09} proposed that a large fraction of the
unmodelled phase variability could be ascribed to a correlation
between the phases and the X-ray flux. This would possibly reflect a
dependence of the hot-spot location on the accretion rate.  To check
this hypothesis in the case of {\src} we substituted the double
integral of the X-ray luminosity in Eq.~\ref{eq:fit} with a linear
term in the X-ray luminosity, $k(t)=D\:L_X(t)$. The value of the
chi-squared we obtain by fitting this model, $\chi^2_r=35.8$, clearly
indicates that the observed phase variability cannot be explained in
terms of such a correlation alone. Nevertheless a phase-flux
correlation may still explain a small fraction of the phase
variability, in addition to the accretion rate dependent torque. By
adding $k(t)$ to Eq.~\ref{eq:fit} the chi-squared of the fit decreases
in fact by $\Delta\chi^2=70$, with one degree of freedom less. The
best-fitting parameters we obtain after the addition of this linear
term to the model are, $\nu(T_0)=11.044884842(6)$ Hz,
$\dot{\nu}(\bar{t})=2.602(6)\times10^{-12}$ Hz s$^{-1}$,
$\beta=0.97\pm0.04$ and $D=0.007(1)$ cycles s erg$^{-1}$.

\section{Discussion}
\label{sec:discussion}

\subsection{The energy spectrum}
\label{sec:discspec}

In Sec~\ref{sec:spectrum} we presented an analysis of the spectra
accumulated by PCA on-board {\it RXTE} (2.5--30 keV), by XRT on-board
{\it SWIFT} (1--10 keV) and by the JEM-X (4.5--25 keV) and ISGRI
(20--150 keV) detectors on-board {\it INTEGRAL}. The phase-averaged
continuum spectrum of {\src} is well described by thermal
Comptonization. During the first few days of the outburst the
temperature of the Comptonizing medium is $kT_e\simgt10$ keV. It then
decreases to values $\approx$ 3--3.5 keV as the source bolometric
luminosity exceeds $\approx3\times10^{37}$ d$_{5.9}^2$ erg s$^{-1}$,
on MJD $\sim$ 55483, without showing further strong deviations from
these values. Correspondingly, the optical depth increases from
$\tau\approx\:$5 to 10. Unfortunately, because there is no
observational coverage of the later stages of the outburst, it is not
possible to ascertain the presence of a corresponding soft-to-hard
transition and at which luminosity does it happen.  The spectral
transition observed during the outburst rise, as well as the softness
of the spectrum characterising the emission of {\src} during most of
the observations, is not typical of faster accreting pulsars. If
compared with {\src}, AMSP attain lower peak luminosities ($\simlt$
few $\times10^{36}$ erg s$^{-1}$), and their spectrum is generally
described by Comptonization in a hotter and more optically transparent
medium (with an electron temperature of $\approx$20-60 keV and optical
depth $\approx$0.7-2.5, see \citealt[][and references
  therein]{Pou06}), not showing strong spectral variability during an
outbursts. Only during the earliest observations, when the source was
fainter than $\approx3\times10^{37}$ d$_{5.9}^2$ erg s$^{-1}$, the
spectrum observed from {\src} somewhat resembles those displayed by
AMSPs. At larger luminosities instead, the observed spectrum is much
more similar to those shown by brighter, non-pulsing, accretors
\citep[see, e.g.,][]{barret02}.  The decrease of the electron
temperature of the Comptonized spectrum as the accretion luminosity
increases can be interpreted in terms of an increase of the flux of
soft photons which efficiently cool the Comptonizing region. In this
context, it is worthwhile to note that based on the analysis of its
correlated colour and aperiodic timing variability, \citet{Alt10} and
\citet{ChkM11} concluded that this source transited from an Atoll-like
to a Z-source like behaviour on MJD 55485, when the X-ray luminosity
was close to attain its peak value.  The hard to soft spectral
evolution reported here is then part of a more general change of the
properties of the accretion flow near the NS surface as the mass
accretion rate increases.

\subsection{The pulse profile shape}

We presented in Sec.~\ref{sec:pulse} an analysis of the 11 Hz pulses
shown by {\src}, based on PCA and XRT observations. The pulse shape
can be modelled by using no more than two harmonic components. A major
change of the pulse properties is observed at MJD $\sim$ 55483, while
the source flux is increasing and the spectrum softening; the
fractional amplitude of the first harmonic component suddenly
decreases from $\sim27\%$ to $\sim2\%$, attaining a value of the same
order of that of the second harmonic (see Fig.~\ref{fig:ampl}). The
drop in the pulse strength is higher at soft energies, since the pulse
amplitude becomes strongly energy dependent, showing a marked increase
of the amplitude with the energy of the photons up to $30$ keV (see
Fig.~\ref{fig:rms_ampl} and \ref{fig:4profiles}).

 The shape of the pulses produced from the surface of an accreting
 rotating NS was modelled by a number of authors \citep[see,
   e.g.,][and references therein]{Mun02,PtnBlb06,Cad07,Lmb09} and
 depends on the geometry of the system (i.e., on the relative angles
 between the spin axis, the magnetic axis and the line of sight), on
 structural properties of the NS (mass, radius, shape and spin), on
 the beaming pattern of the emitted radiation, as well as on the size,
 temperature and brightness contrast of the accretion cap with respect
 to the rest of the NS surface.  During most of the outburst the
 pulses shown by {\src} are of low amplitude (few per cent) and
 simple, similar to the majority of AMSP (see \citealt[][L09
   hereafter]{Lmb09}). However, the amplitude decrease by more than a
 factor of 10 taking place on MJD $\sim$ 55483 on a timescale of less
 than one day (see Fig.~\ref{fig:ampl}), indicates how some of the
 properties that determine the pulse amplitude must have changed
 significantly as the X-ray luminosity was increasing during the
 outburst rise. Variations of the pulse amplitude of a similar amount
 were already observed from a number of AMXPs; swings from 3 to 25 per
 cent have been observed from XTE J1807--294 \citep{Zhg06}, and from 4
 to 14 per cent in the case of SAX J1808.4--3658 \citep{Hrt08}. In
 those cases, though, the amplitude variations were closely related to
 flaring episodes during the outbursts. On the other hand, the
 decrease of the pulsed fraction shown by {\src} is step-like and no
 significant correlation holds between the pulsed fraction and the
 X-ray flux afterwards. In the following a number of hypotheses to
 account for the observed variation of the pulse properties during the
 outburst rise is discussed. Unfortunately, a coverage of the later
 stages of the outburst decays is not available, preventing to
 investigate if a similar transition happens as the source dims and at which
 luminosity would it take place.

A change of the spot size is hardly the main driver of the observed
pulse fraction decrease; the spot size is in fact expected not to vary
by more than $\sim 10\%$ as the luminosity doubles, and the results
obtained by \citet{Mun02} and L09 show how the dependence of the pulse
amplitude on the spot size is much weaker than it would be needed to
explain the observed variation.

Pulse amplitude variations can be explained in terms of movements of
the hot spot location on the NS surface, possibly caused by the
in-falling matter getting attached to a different set of magnetic
field lines. An increase of the spot latitude can be particularly
efficient in decreasing the pulse amplitude, especially if the angle
between the magnetic and the spin axis is small. This was invoked by
L09 to explain some of the properties of the pulses shown by AMSPs,
such as small and variable amplitudes, and unstable pulse
phases. According to their results a change of roughly $15-30^{\circ}$
of the spot latitude towards the magnetic pole would be able to produce an
amplitude decrease of the order of that observed from {\src}. Further,
under the hypothesis that the two spots on the NS surface are
antipodal, a re-sizing, or shift, of the impact region of the
in-falling matter on the NS could increase the fraction of rotational
phases during which both spots on the NS surface become visible. This
would indeed contribute to explain the observed reduction of the pulse
amplitude and strengthening of the second harmonic amplitude relative
to the first one. Although this scenario is then in principle able to
explain the decrease of the pulse amplitude integrated over the entire
PCA energy-band, it gives no sufficient clues for the simultaneous
change of the energy dependence of the pulse profile shape.

An amplitude variation of the order of that observed could also be
produced by a significant change of the angular distribution of the
photons that reach the observer, possibly triggered by the source
luminosity crossing a certain threshold level \citep[see, e.g., ][who
  interpreted as such the luminosity dependent changes of the pulse
  profile of the 42 s high-field pulsar, EXO
  2030+375]{parmar89b}. While the radiation transmitted through a slab
is strongly peaked along the normal to the NS surface, scattered
radiation is expected to peak at intermediate angles, in a fan-like
emission pattern \citep[e.g.][]{SunTtr85}. The relations given by
\citet[][see, e.g., Eq.~51 and 61]{PtnBlb06} show that an amplitude
change of the same order of that observed can be reproduced if the
angular distribution of the emitted photons switches from being
strongly peaked along the normal [$I(\alpha)\propto(1+h\cos{\alpha})$,
  where $\alpha$ is the angle to the normal and $h\simeq$ 2], to a fan
like distribution (h$\simeq$ -1). However, such a drastic change of
the pattern of the emitted radiation would produce a change of
$\sim0.5$ in the pulse phase, which is not observed.

The decrease of the pulse amplitude shown by {\src}, as well as its
spectral hardening, take place roughly simultaneously to the spectral
softening of the phase-averaged emission discussed in the previous
section.  It is then intriguing to explain these transitions in terms
of a variation of the properties of the matter flow close to the NS
surface. The observed behaviour would be in fact explained if a large
fraction of the source luminosity, and in particular its softer part,
becomes not pulsed during the outburst rise.  This could happen
because the inner rim of the disc approaches the NS surface, with an
increased part of the disc non-pulsed emission falling in the
considered energy band, and/or because an increasing fraction of the
in-falling matter is not channelled to the magnetic poles, but rather
accretes more evenly at the NS equator in a sort of optically thick
boundary layer. The properties of the phase-averaged spectrum
presented in Sec.~\ref{sec:spectrum} would favour the latter
hypothesis. The possibility that not all the plasma from the accretion
disc is lifted off by magnetic field lines at the magnetospheric
boundary, with a fraction penetrating between the magnetic field lines
via a Rayleigh-Taylor instability or because of incomplete coupling of
the plasma and the field lines was studied by, e.g., \citet{SprTmm90}
and \citet[][see references therein, also]{Mll98}. Recent 3D MHD
simulations of disc accretion onto a magnetised NS performed by
\citet{Rmn08} showed how a transition to this twofold accretion
channel may take place when the mass accretion rate becomes larger
than $\simeq 0.1 (B/10^9\mbox{G})^2\dot{M}_{Edd}$. Above this level it
could then be that only a fraction $\sim 10$\% of the accreted mass is
channelled to the poles. The rest would be accreted preferentially at
the star equatorial belt producing the observed optically thick
Comptonized component, reminiscent of emission from non pulsing NS in
LMXB. The hardening of the energy dependence of pulse amplitude would
be naturally explained by this scenario. The pulsed emission
originating from the polar caps is in fact expected to be harder than
the non-pulsed emission produced in a denser equatorial boundary
layer. Interestingly, an increase of the pulse amplitude with energy
similar to that of {\src} was observed only from three AMSPs, SWIFT
J1756.9-2508 \citep{Ptr10}, Aql X-1 \citep{Csl08} and SAX J1748.9-2021
\citep{Ptr09}, the last two being sources that showed pulsations only
intermittently. To conclude, it has to be noted that a pulse amplitude
increase at energies $\simgt 50$ keV was also observed by
\citet{FlnTtr07} from IGR J00291+5934. Since it concerned high photon
energies, the authors interpreted such an increase as due to the
decrease of the electron scattering cross section with energy. This
effect clearly does not explain the pulse amplitude properties
observed from {\src}, as well as from the other mentioned sources,
since the pulse amplitude is observed to increase even at low ($\simgt
2.5$ keV) energies.

\subsection{Phase evolution.}
\label{sec:disc_torque}
We presented in Sec.~\ref{sec:timing} and \ref{sec:torque} an analysis
of the temporal evolution of the phase of the pulse shown by {\src}
during its 2010 outburst, based on the whole data set available of PCA
and XRT observations. The results obtained confirm the earlier finding
that the NS in this system accelerates while it accretes mass
(P11). While the analysis performed on the phases computed over the
first and second harmonic gives slightly different values of the
average frequency derivative, a value of
$<\dot{\nu}>=1.48(2)\times10^{-12}$ Hz $s^{-1}$ encompasses both
estimates.

 The observed frequency derivative of the signal emitted by a pulsar
 is the sum of the intrinsic change of the NS spin and of the
 acceleration of the pulsar with respect to the observer. While the
 apparent spin frequency derivative due to centrifugal acceleration
 ($\simlt\nu \mu^2 d/c\simeq1.4\times10^{-18}$ d$_{5.9}$ Hz s$^{-1}$,
 \citealt{shklovskii70}, where $\mu=85\pm10$ km s$^{-1}$ is the
 velocity of the proper motion of Terzan 5 estimated by
 \citealt{Fer09}) and to galactic acceleration
 ($\simeq2\times10^{-17}$ Hz s$^{-1}$, obtained by inserting the
 galactic coordinates of {\src} in the relation given by
 \citealt{damour91}) can be safely neglected, the acceleration of the
 pulsar in the gravitational field of the cluster can in principle
 have a larger effect. \citet{phinney93} estimated the maximum
 acceleration along the line of sight for an object located at a
 projected angular distance $\theta_{\bot}$ from the cluster centre as
 $1.1\:\nu\:G\:M_{cyl}(<\theta_{\bot})/(c\:\pi\:
 d^2\:\theta_{\bot}^2)$ Hz s$^{-1}$, where $M_{cyl}(<\theta_{\bot})$
 is the mass within a projected distance smaller than
 $d\:\theta_{\bot}$ from the cluster centre. Considering the cluster
 parameters estimated by \citet{Lnz10} from the surface brightness
 profile of Terzan 5 (core and tidal radius of $9''$ and $277''$,
 respectively, and a total bolometric luminosity of $8\times10^5$
 L$_{\sun}$) and the projected distance of the pulsar from the cluster centre
 ($\bar{\theta}_{\bot}\simeq4''$) to integrate the King density
 profile \citep{king62} up to $\theta_{\bot}$, yields  a mass
 $M_{cyl}(<\bar{\theta}_{\bot})\simeq5\times10^4$ M$\sun$, where a
 mass-to-light ratio of 3 was assumed. Inserting such a value in the
 relation given above yields an apparent frequency derivative of
 $\simeq6\times10^{-15}$ Hz s$^{-1}$, which is more than two orders of
 magnitude lower than the observed derivative and can be safely
 neglected.

The average derivative of the signal frequency is  compatible
with the average acceleration imparted to the NS by the accretion of the
Keplerian angular momentum of disc matter. By considering only this
torque, $N_0$, and assuming that the X-ray luminosity tracks the mass
accretion rate (see Sec.~\ref{sec:torque}), the expected spin
frequency derivative of an accreting NS is
\begin{equation}
\dot{\nu}=\frac{N_0}{2\pi I_*}=\frac{\sqrt{GM_*}}{2\pi I_*} \dot{M} \sqrt{R_{in}}=\frac{ R_*}{2\pi I_*\eta\sqrt{GM_*}}  \sqrt{R_{in}} \; L_X,
\label{eq:disc_torque}
\end{equation} 
where $R_{in}$ is the inner disc radius. By inserting the measured
value of $<\!\dot{\nu}\!>$ in Eq.~\ref{eq:disc_torque}, an estimate of the
average inner disc size of $<\!R_{in}\!>\simeq64 \;\eta^{2}\;R_{*,10}^{-2}\;
M_{*,1.4}\; I_{45}^{2} (<\!L_{37}\!>/\:5)^{-2}$ km is obtained, where
$R_{*,10}$ is the NS radius in units of 10 km, $M_{*,1.4}$ the NS mass
in units of 1.4 M$_\odot$, $I_{*,45}$ the NS moment of inertia in units
of $10^{45}$ g cm$^2$ and $L_{37}$ is the accretion luminosity in units
of $10^{37}$ erg s$^{-1}$ . Such a value is clearly compatible with
the upper and lower bounds set by the presence of pulsations, $R_*<
R_{in}\simlt R_{C}$, where $R_C=338$ $M_{*,1.4}^{1/3}$ km is the
corotation radius around the NS in {\src} (P11).

The analysis presented in Sec.~\ref{sec:timing} shows that a constant
frequency model is not compatible with the observed evolution of the
phases of {\src}. Yet, values of the reduced $\chi^2$ in excess of
one are obtained even if an 8-th order polynomial expansion is
considered as a model, indicating the presence of timing noise. Before
discussing the results obtained by modelling the observed phase
evolution with a physically-motivated torque model, the influence of
such a noise on the phase determination has to be discussed.

Noisy variability in the phases of accreting pulsars is not new and
manifests itself through the presence of unmodelled phase residuals
with respect to the considered timing solution. In particular, it
affects the pulses shown by AMSPs and it was subject of a profound
analysis and discussion \citep[see, e.g.,][and references
  therein]{Brd06, papitto07, H08, Rig08, Lmb09,Ptr09}. The analysis of the
temporal evolution of the phases of the pulse profile shown by a
pulsar allows in principle the most accurate possible knowledge of the
rotational evolution of the NS. However, irregular variability of the
pulse phases not related to intrinsic spin changes of the NS hinders
this possibility. Indeed, this is particularly troublesome for
transiently accreting pulsars such as AMSP which are expected to
experience relatively small frequency variations during their
outbursts. In those cases the amplitude of the phase variability
induced by timing noise can be comparable to that produced by a spin
frequency derivative. This is certainly not the case of {\src}, since,
despite the presence of timing noise, the pulse phase is observed to
perform more than one phase cycle across the outburst (see
Fig.~\ref{fig:phase1} and \ref{fig:phase2}).  This source can be
therefore regarded as a particularly relevant case to the study of
accreting pulsars, since the the contributions of timing noise and of
the rotational evolution to the pulse phase variations may be
disentangled to some extent.

An evident property of the phase noise of {\src} is that affects
roughly three times more weakly the phases computed over the second
harmonic than those of the fundamental. Together with the much easier
description of the evolution of the second harmonic phases in terms of
a physically plausible torque model (see Sec.~\ref{sec:torque} and in
the following), this provides an observational indication of how the
second harmonic component might be a better tracer of the spin
frequency evolution of an accreting pulsar.  The enhanced stability of
the second harmonic component with respect to the fundamental was
already observed in a number of AMSPs. Sudden jumps of the first
harmonic phases not accompanied by a corresponding variation of the
second harmonic were observed during a couple of outbursts of SAX
J1808.4--3658 \citep{Brd06,Hrt09}, while during other outbursts a
comparable amount of noise affected the two harmonic components
\citep{H08}. A similar phenomenology is also observed from SWIFT
J1749.4--2807 (Papitto et al., in prep.). Further, in a couple of
cases (XTE J1807--295, \citealt{Rgg08}; IGR J17511-3057,
\citealt{Rgg11}) the noise affecting the second harmonic component was
found to be significantly milder than that affecting the fundamental
frequency; moreover, only by modelling the second harmonic component
 a physically plausible timing solution could be proposed for
these systems. While the different stability of the various harmonic
components was taken into account by \citet{H08} to derive a minimum
variance estimator for the phases of the pulse profile, not many
attempts have been made to explain such a different behaviour of the
phases computed over different Fourier components. If the observed
variability of the pulse amplitudes and phases is due to irregular
motion of the spot on the NS surface (L09) similar shifts should be
observed in all the Fourier components. This has been observed and
interpreted as such only in the case of XTE J1814-338
\citep{papitto07,watts08}. While it could well be that substantial
changes in the spot-shape or in the radiation-pattern of the emitted
radiation determine a change of the pulse shape, this does not seem to
naturally explain the enhanced stability of the second harmonic
observed in a number of cases, and why only the use of the second
harmonic phases as a tracker of the NS spin frequency generally
produces results in accordance with theoretical expectations.

A simple model proposed by \citet{riggio11} explains a similar
behaviour in terms of modest variations of the relative intensity
received by the two polar caps onto the NS surface, if the geometrical
properties of the system make the signal coming from the two spots of
similar amplitude. It is assumed that the pulse profile observed from
an accreting pulsar is given by the sum of two signals with a similar
harmonic content, emitted by two nearly antipodal spots on the NS
surface ($\Delta\lambda\approx \pi$, $\Delta \gamma\approx 2 \gamma$,
where $\gamma$ and $\lambda$ are the latitude and longitude of the
primary spot, respectively, and $\Delta\lambda$ and $\Delta\gamma$ are
the differences between the coordinates of the two spots), as it is
the case if the magnetic field has a dipolar shape. The sum of the two
signals (the total profile) will be the result of a destructive
interference for what concerns the fundamental harmonic of the signal,
since it is the sum of two signals with a phase difference of
$\approx\pi$. A constructive interference develops instead for the
second harmonic of the total profile, since it is the sum of two
signals with the same phase. The destructive interference regarding
the fundamental frequency leads to (i) small values of the pulsed
fraction of the first harmonic component, possibly comparable to that
of the second harmonic, and (ii) large swings of the phase of the
fundamental of the total profile due to modest variations of the
relative intensity of the signals emitted by the two caps. The
magnitude of these effects is the largest when the spots on the NS
surface are exactly antipodal and produce signals that are observed at
a similar amplitude. In this case swings up to 0.5 phase cycles can be
shown by the phase computed over the fundamental frequency of the
observed profile, without  correspondingly large variations of the
phase of the second harmonic.  The maximum variability of the phase of
the first harmonic component, as well as the maximum reduction of its
amplitude, is obtained when the intensity of the signals observed
from the two magnetic caps is very similar. This happens if the
system is viewed nearly edge on and/or if the rotator is nearly
orthogonal, that is with an angle between the spin and the magnetic
axes of $\approx 90^{\circ}$. Under these assumptions such a simple
model can reproduce phase movements of the order of those observed
from {\src}, as well as the enhanced stability observed from the
second harmonic phases, in terms of variations of few per cent of the
ratio between the intensity observed by the two spots. A thorough
quantitative discussion of how this model applies to this, and other
cases is deferred to a dedicated paper (Riggio et al., in prep.).

In addition to the smaller residuals with respect to phenomenological
models, an enhanced stability of the second harmonic phases is also
indicated by the possibility of describing their behaviour by using a
physically plausible model. In Sec.~\ref{sec:timing} and
\ref{sec:torque} we showed how the description of the pulse phases
determined using the second harmonic component is significantly
improved by assuming that the frequency derivative of the signal
depends on a power close to one of the X-ray flux. The
descritption of the pulse phases is further improved by the assumption of
a phase-flux correlation such as that proposed by \citet{Ptr09}, even
if by a much lesser extent. On the other hand, the spin frequency
derivative evaluated by considering the first harmonic phases shows
oscillations that are not correlated with the X-ray luminosity.

Under the hypothesis that the second harmonic phases are a good
tracker of the NS spin, we can infer some of the properties on the
size of the inner radius of the accretion disc and on its dependence
on the X-ray luminosity. To this end we consider the results presented
in Sec.~\ref{sec:torque}, where the phases of the second harmonic were
fitted by assumung a relation, $\dot{\nu}\propto L_X^{\beta}$, for the
dependency of the spin-up torque on the accretion luminosity.
According to standard accretion disc theory, the angular momemtum is
transferred from the disc matter to the magnetosphere at the inner
disc radius $R_{in}$. In turn, the size of the inner edge of the disc
is determined by evaluating where the stresses exerted by the magnetic
field lines on the accreting plasma become dominant with respect to
the viscous stresses, that grant angular momentum redistribution in the
disc far from the magnetosphere \citep[see, e.g.,][and references
  therein]{Ghs07}. The location of the inner disc radius is usually
expressed as a fraction $\xi$ of the Alfven radius evaluated in the
case of spherical symmetric accretion
\begin{equation}
R_{in}=71 \:\xi\: \eta^{2/7}\: M_{*,1.4}^{1/7}\: R_{*,10}^{-2/7}\:L_{37}^{-2/7}\: \mu_{27}^{4/7}\: \mbox{km}.
\label{eq:rin}
\end{equation} 
Here, $\mu_{27}$ is the magnetic-dipole moment in units of $10^{27}$ G
cm$^3$, and the factor $\xi$ is determined by capability of the field
lines to thread the disc matter, and by the magnetic pitch, which in
turn is limited by the reconnection-mechanisms of the field lines
through the disc (\citealt{GL79,Wng96}, who quote values of $\xi$
between 0.5 and 1, respectively; see also the discussion of these
models given by \citealt{bozzo09}).  Inserting this expression in
Eq.~\ref{eq:disc_torque}, it can be seen that the spin frequency
derivative is expected to scale as $\dot{\nu}\propto L_X^{6/7}$, if
the only relevant torque is that exerted by matter orbiting in a
Keplerian disc.

 The dependence on the X-ray luminosity of the spin frequency
 derivative evaluated from the phases of the second harmonic shown by
 {\src}, $\beta=1.07\pm0.03$, is steeper than what is predicted by
 standard accretion theory. Values of $\beta$ close to unity were already
 reported by \citet{bildsten97} for the 0.47 s accreting pulsar in a
 LMXB, GRO J1744--28 ($\beta=0.957\pm0.026$) and the 105 s high field
 pulsar, A0535+26 ($\beta=0.951\pm0.026$), as well as by
 \citet{parmar89} for the 42 s high field accreting pulsar, EXO
 2030+375 ($\beta=1.21\pm0.13$). Expressing the dependence of the
 inner disc radius on the accretion luminosity as $R_{in}\propto
 L_X^{-\alpha}$, the estimate of $\beta$ found for {\src} translates
 in a value of $\alpha=2(1-\beta)=-0.14\pm0.06$. A negative value of
 $\alpha$ would imply a positive correlation of the inner disc radius
 with the X-ray luminosity, which seems unlikely on physical
 grounds. On the other hand, a value close to zero indicates a very
 weak, if any, dependence of the inner disc radius on the accretion
 luminosity.  A similar result, $\beta=0.97(4)$ [$\alpha=0.06(8)$], is
 obtained when adding to the model used to fit the phases a term
 depending linearly on the X-ray luminosity.

By considering the value of the spin up torque,
$\dot{\nu}(\bar{t})=2.755(5)\times10^{-12}$ Hz s$^{-1}$, attained when
the X-ray luminosity reached its peak value,
$L_{X}(\bar{t})=9.23(4)\times10^{37}$ d$_{5.9}^2$ erg s$^{-1}$, and
assuming that the NS accretes the Keplerian angular momentum of matter
at the inner disc boundary $R_{in}$ (Eq.~\ref{eq:disc_torque}), the
inner edge of the disc is estimated as
$R_{in}(\bar{t})=65(1)\:\eta^{2}
I_{*,45}^2\:M_{*,1.4}\:R_{*,6}^{-2}\:d_{5.9}^{-4}$ km.  A slightly
smaller estimate, $\simeq58$ km, is obtained if the value of the
frequency derivative obtained after adding a linear term in the X-ray
luminosity to the model used to fit the phases,
$\dot{\nu}(\bar{t})=2.602(6)\times10^{-12}$ Hz s$^{-1}$, is
considered. The error on the distance ($\simeq8\%$ of its estimate) is
the largest source of uncertainty on the disc size measure. By taking
it into account, the estimate of $R_{in}(\bar{t})$ lies between 47 and
93 km (1$\sigma$ confidence level).  Inserting the measured values of
$R_{in}$ and $L_{max}$ in Eq.~\ref{eq:rin} yields to an estimate of
the magnetic-dipole moment, $\mu_{27}=2.6(2)\: \xi^{-7/4}\:
d_{5.9}^{-6}\: I_{*,45}^{7/2}\: R_{*,10}^{-3}\: M_{*,1.4}^{3/2}\:
\eta^{-15/4}$, which has a similarly large uncertainty and for
$\xi\simeq0.5$ varies between 5 and 16. A magnetic surface flux
density between $\sim 5 \times 10^{9}$ and $\sim 1.5 \times10^{10}$ G is
therefore likely.

Since {\src} is a relatively slow pulsar, a coupling between magnetic
field lines and plasma rotating in the accretion disc beyond the inner
disc radius transfers additional angular momentum to the NS. To take
into account this effect, the torque on the NS can be expressed as
\begin{equation}
N\approx1.4\:N_0\:\frac{1-\omega_S/\omega_C}{1-\omega_S}
\label{eq:magtorque}
\end{equation}\citep{GL79}. Here,
$\omega_S=2\pi\nu\sqrt{R_{in}^3/GM}$ is the fastness parameter while
$\omega_C$ is the critical fastness, which takes values between 0.35
and 1 depending on the description of the interaction between the
field and the disc \citep{GL79,Wng96}.  Taking the value of
$\omega_C=0.6$ obtained by \citet{Rom02} from numerical simulations,
and equating the observed torque to Eq.~\ref{eq:magtorque}, the
estimate of the inner disc radius at $\bar{t}$ reduces to roughly half
of the estimate given above, $R_{in}\simeq36\: d_{5.9}^{-4}$ km, where
only the steep dependence on the distance was made explicit. The
modelling of the phases of the second harmonic with a phisically
plausible torque model yields therefore a value of the inner disc
radius compatible with the estimate given by \citet{Mll11} on a
spectroscopic basis, $41(5)\: M_{*,1.4}$ km; the latter value was obtained
by interpreting in terms of relativistic broadening the profile of the
iron K-$\alpha$ line detected in the spectrum of the source observed
by {\it Chandra} on 55493.445, (13.632 d since $T_0$), when the X-ray
luminosity was a factor $\approx0.6$ lower than at the peak. Still,
these estimates appear slightly larger than the inner disc radius
implied by the observation of a QPO at 815 Hz during the outburst peak
\citep{Alt10}, $R_{in}\simeq 19 \: M_{*,1.4}$ km, if such a feature is
interpreted in terms of a Keplerian orbital frequency.  Given the
steep dependence of our inner disc radius estimate on the
distance to the source, the size given by the Keplerian
interpretation of the QPO would be reconciled with the value indicated
by the torque analysis presented here if a distance of $\simgt 7$
kpc to the Globular cluster is assumed. A similar value is marginally compatible
with the most recent determination ($5.9\pm0.5$ kpc), though it has to
be noted that heavy, differential reddening affecting observations of
this Globular cluster make the determination of its distance somewhat
problematic \citep[][and references therein]{Lnz10}.

\section*{Acknowledgments}

This work is supported by the operating program of Regione Sardegna
(European Social Fund 2007-2013), L.R.7/2007, ``Promoting scientific
research and technological innovation in Sardinia'', by the Italian Space
Agency, ASI-INAF I/009/10/0 contract for High Energy Astrophysics, as
well as by the Initial Training Network ITN 215212: Black Hole
Universe funded by the European Community. AP also acknowledges the
support of the grants AYA2009-07391 and SGR2009-811, as well as of the
Formosa program TW2010005 and iLINK program 2011-0303.

\bibliography{biblio}
\bibliographystyle{mn2e}

\newpage

\onecolumn	

\begin{center}	
\begin{landscape}
\begin{longtable}{lccccccccc}

\caption[l]{Best-fitting parameters of the spectral continuum
  evaluated from the 2.5--30 keV spectra observed by PCA on-board {\it
    RXTE}. Columns are: Observation ID (labels refer to the
    groups of observations defined in text); Start time of the
  observation (MJD since $T_0$ = MJD 55479.81229); End time of the
  observation (MJD since $T_0$); Exposure (ks); $kT_e$ (keV); $\tau$;
  $kT_{S}$ (keV); Unabsorbed flux in the 2.5--25 keV band
  ($10^{-8}$ erg cm$^{-2}$ s$^{-1}$); Bolometric luminosity calculated
  by extrapolating the best-fitting model to the 0.05--150 keV energy
  band ($d_{5.9}^2\:10^{37}$ erg cm$^{-2}$ s$^{-1}$); Reduced
  chi-squared of the fit.\label{tab:xtespectra} The quoted errors are
  evaluated at the 90\% confidence level on each parameter
  ($\Delta\chi^2=2.706$).} \\ \endfirsthead \multicolumn{10}{c}

{{\tablename\	\thetable{}	--	continued	from	previous	page}}	\\				

\hline \hline
 ObsID & T$_{start}$ (MJD) & T$_{end}$ (MJD) &
Expo (ks) &  kT$_e$ (keV) & $\tau$ & kT$_{S}$ (keV) &
F($10^{-8}$ erg cm$^2$ s$^{-1}$) & L ($d_{5.9}^2\:10^{37}$ erg
s$^{-1}$)&$\chi^2$/dof \\

\hline																											
\hline

\endhead																											
\hline	\multicolumn{10}{c}{{Continued	on	next	page}}	\\																						
\endfoot																											
\hline																											
\endlastfoot																											
\hline																						
 ObsID & T$_{start}$ (MJD) & T$_{end}$ (MJD) &
Expo (ks) &  kT$_e$ (keV) & $\tau$ & kT$_{S}$ (keV) &
F($10^{-8}$ erg cm$^2$ s$^{-1}$) & L ($d_{5.9}^2\:10^{37}$ erg
s$^{-1}$)&$\chi^2$/dof \\
\hline																													
\hline							
				
95437-01-01-00 -- Obs. [B]   & 2.1964&   2.2328  &2.3&$6.7_{-0.2  }^{+0.2  }  $&$8.22_{-0.09 }^{+0.07}$ &  $0.80_{-0.04 }^{+0.04 }$  & 0.340(2) &1.91(4) & 65.1/49  \\    
95437-01-02-00  & 3.3740&   3.5506  &7.5&$3.7 _{-0.03 }^{+0.04 }  $&$9.57_{-0.03 }^{+0.06}$ &  $0.840_{-0.006}^{+0.03 }$ & 0.966(2) &4.81(2) & 51.7/49  \\   
95437-01-02-01 -- Obs. [C]  & 3.9028&   4.0665  &8.2&$3.62_{-0.03 }^{+0.01 }  $&$9.43_{-0.10 }^{+0.02}$ &  $0.85_{-0.05 }^{+0.01 }$  & 1.033(2) &5.14(2) & 57.9/49  \\    
95437-01-03-00  & 4.6214&   4.6506  &1.4&$3.36_{-0.02 }^{+0.02 }  $&$9.94_{-0.02 }^{+0.02}$ &  $0.81_{-0.01 }^{+0.01 }$  & 1.122(3) &5.62(3) & 46.7/49  \\    
95437-01-03-01  & 4.7560&   4.7895  &1.4&$3.34_{-0.03 }^{+0.02 }  $&$9.78_{-0.02 }^{+0.02}$ &  $0.82_{-0.01 }^{+0.02 }$  & 1.142(3) &5.73(3) & 50.9/49  \\    
95437-01-03-02  & 4.8162&   4.8549  &2.4&$3.34_{-0.04 }^{+0.02 }  $&$9.62_{-0.04 }^{+0.12}$ &  $0.83_{-0.02 }^{+0.05 }$  & 1.170(3) &5.87(4) & 58.0/49  \\    
95437-01-03-03  & 4.9432&   5.1165  &8.0&$3.33_{-0.01 }^{+0.01 }  $&$9.39_{-0.01 }^{+0.01}$ &  $0.828_{-0.006}^{+0.005}$ & 1.201(2) &6.06(2) & 61.5/49  \\   
95437-01-04-00  & 5.6019&   5.7049  &3.5&$3.22_{-0.06 }^{+0.01 }  $&$8.82_{-0.06 }^{+0.04}$ &  $0.88_{-0.07 }^{+0.02 }$  & 1.404(3) &7.11(3) & 57.6/49  \\    
95437-01-04-01  & 5.7347&   5.9014  &6.1&$3.18_{-0.04 }^{+0.04 }  $&$8.97_{-0.06 }^{+0.09}$ &  $0.86_{-0.02 }^{+0.03 }$  & 1.473(3) &7.49(4) & 60.1/49  \\    
95437-01-05-00  & 6.6486&   6.7514  &4.2&$2.97_{-0.01 }^{+0.01 }  $&$8.31_{-0.07 }^{+0.02}$ &  $0.90_{-0.02 }^{+0.01 }$  & 1.613(3) &8.29(3) & 62.9/49  \\    
95437-01-05-01  & 6.7743&   6.9471  &8.9&$2.905_{-0.008}^{+0.007} $&$8.56_{-0.01 }^{+0.02}$ &  $0.862_{-0.004}^{+0.004}$ & 1.528(3) &7.94(3) & 66.6/49  \\   
95437-01-06-000 & 7.4932&   7.7473  &8.8&$2.938_{-0.008}^{+0.008} $&$8.29_{-0.01} ^{+0.01}$ &  $0.853_{-0.004}^{+0.004}$ & 1.658(3) &8.68(3) & 47.5/49  \\   
95437-01-06-00 -- Obs. [D] & 7.7527&   7.8804  &4.7&$3.001_{-0.01 }^{+0.009} $&$8.69_{-0.01 }^{+0.02}$ &  $0.829_{-0.005}^{+0.004}$ & 1.771(3) &9.23(4) & 55.3/49  \\   
95437-01-07-00  & 8.3375&   8.5604  &7.0&$3.08_{-0.2  }^{+0.02 }  $&$8.35_{-0.10 }^{+0.05}$ &  $0.84_{-0.02 }^{+0.02 }$  & 1.704(3) &8.89(4) & 32.1/49  \\    
95437-01-07-01  & 8.9969&   9.0577  &2.3&$2.73_{-0.02 }^{+0.06 }  $&$7.88_{-0.07 }^{+0.10}$ &  $0.88_{-0.02 }^{+0.02 }$  & 1.334(3) &7.06(3) & 57.0/49  \\    
95437-01-08-00  & 9.7167&   9.8493  &5.2&$2.93_{-0.02 }^{+0.009}  $&$7.97_{-0.02 }^{+0.01}$ &  $0.886_{-0.004}^{+0.003}$ & 1.348(3) &7.02(3) & 42.8/49  \\   
95437-01-10-01  & 12.3310&  12.4341 &4.5&$3.18_{-0.01 }^{+0.01 }  $&$8.86_{-0.01 }^{+0.01}$ &  $0.799_{-0.005}^{+0.005}$ & 1.214(2) &6.30(3) & 47.2/49  \\   
95437-01-10-02  & 13.4540&  13.5523 &4.2&$3.05_{-0.4  }^{+0.08 }  $&$8.3_{-0.4  }^{+0.2 }$  &  $0.97_{-0.05 }^{+0.01 }$  & 1.123(3) &5.63(5) & 52.1/47  \\     
95437-01-10-03  & 14.4310&  14.5554 &4.3&$2.84_{-0.01 }^{+0.07 }  $&$9.95_{-0.01 }^{+0.07}$ &  $0.916_{-0.010}^{+0.004}$ & 1.129(2) &5.63(2) & 60.7/49  \\   
95437-01-10-04  & 15.5345&  15.5736 &2.4&$3.19_{-0.02 }^{+0.02 }  $&$9.28_{-0.04 }^{+0.02}$ &  $0.81_{-0.01 }^{+0.02 }$  & 1.089(2) &5.57(3) & 65.0/49  \\    
95437-01-10-07  & 15.6012&  15.6582 &2.5&$2.98_{-0.009}^{+0.01 }  $&$9.85_{-0.01 }^{+0.08}$ &  $0.86_{-0.01 }^{+0.04 }$  & 1.102(2) &5.55(3) & 40.4/49  \\    
95437-01-10-05  & 16.5797&  16.7062 &4.8&$2.95_{-0.1  }^{+0.04 }  $&$10.2_{-0.8  }^{+0.1 }$ &  $0.94_{-0.02 }^{+0.03 }$  & 1.107(2) &5.45(2) & 57.7/49  \\    
95437-01-10-06  & 17.6991&  17.8117 &4.7&$2.908_{-0.006}^{+0.006} $&$10.98_{-0.01} ^{+0.01}$&  $0.873_{-0.004}^{+0.002}$ & 1.059(1) &5.23(1) & 80.3/49 \\  
95437-01-11-07  & 18.6880&  18.7110 &1.3&$2.88_{-0.02 }^{+0.01 }  $&$11.41_{-0.80}^{+0.02}$ &  $0.8_{-0.1  }^{+0.1  }$   & 1.016(3) &5.10(3) & 49.4/49  \\     
95437-01-11-00  & 18.7371&  18.7903 &2.8&$3.00_{-0.01 }^{+0.01 }  $&$10.45_{-0.02}^{+0.02}$ &  $0.871_{-0.009}^{+0.006}$ & 1.006(2) &5.00(2) & 55.1/49  \\     
95437-01-11-01  & 19.5358&  19.6465 &4.2&$3.11_{-0.01 }^{+0.01} $&$10.09_{-0.04 }^{+0.03}$&  $0.89_{-0.02 }^{+0.08 }$  & 0.997(2)   &4.93(2) & 53.2/49  \\   
95437-01-11-02  & 20.5158&  20.6291 &4.3&$3.24_{-0.02 }^{+0.01 }  $&$10.01_{-0.06 }^{+0.02}$&  $0.88_{-0.02 }^{+0.02 }$  & 0.993(2) &4.92(2) & 54.4/49  \\   
95437-01-11-03  & 21.4310&  21.5443 &4.2&$3.35_{-0.01 }^{+0.02 }  $&$9.62_{-0.02 }^{+0.02}$ &  $0.885_{-0.006}^{+0.006}$ & 0.990(2) &4.90(2) & 67.2/49  \\   
95437-01-11-04  & 22.5923&  22.6971 &5.4&$3.48_{-0.08 }^{+0.01 }  $&$9.09_{-0.09}^{+0.02}$  &  $0.87_{-0.02 }^{+0.01 }$  & 0.921(2) &4.59(2) & 79.7/49  \\     
95437-01-11-05  & 23.3173&  23.3447 &2.1&$3.32_{-0.03 }^{+0.02 }  $&$10.03_{-0.03 }^{+0.02}$&  $0.84_{-0.04 }^{+0.02 }$  & 0.952(2) &4.73(3) & 39.7/49  \\   
95437-01-11-08  & 23.3817&  23.4403 &2.4&$3.34_{-0.07 }^{+0.03 }  $&$9.75_{-0.22}^{+0.09}$  &  $0.86_{-0.02 }^{+0.02 }$  & 0.942(2) &4.68(3) & 72.8/49  \\     
95437-01-11-06  & 24.3567&  24.4812 &5.4&$3.39_{-0.04 }^{+0.01 }  $&$9.66_{-0.05 }^{+0.06}$ &  $0.88_{-0.01 }^{+0.03 }$  & 0.939(2) &4.64(2) & 54.5/49  \\    
95437-01-12-00  & 25.4164&  25.5256 &4.4&$3.36_{-0.01 }^{+0.01 }  $&$9.88_{-0.02 }^{+0.02}$ &  $0.876_{-0.006}^{+0.006}$ & 0.931(2) &4.60(2) & 61.3/49  \\   
95437-01-12-01  & 26.4454&  26.5701 &5.6&$3.48_{-0.02 }^{+0.02 }  $&$9.37_{-0.05 }^{+0.03}$ &  $0.91_{-0.03 }^{+0.02 }$  & 0.916(2) &4.51(2) & 80.8/49  \\    
95437-01-12-03  & 28.3430&  28.4708 &5.7&$3.45_{-0.05 }^{+0.03 }  $&$9.51_{-0.11}^{+0.07}$  &  $0.92_{-0.05 }^{+0.02 }$  & 0.880(2) &4.32(2) & 62.9/49  \\     
95437-01-12-04  & 29.3238&  29.4527 &5.6&$3.38_{-0.01 }^{+0.01 }  $&$10.10_{-0.01 }^{+0.02}$&  $0.851_{-0.007}^{+0.010}$ & 0.854(2) &4.23(2) & 60.8/49  \\  
95437-01-12-05  & 30.3362&  30.4527 &4.7&$3.39_{-0.01 }^{+0.04 }  $&$9.76_{-0.02 }^{+0.03}$ &  $0.854_{-0.005}^{+0.030}$ & 0.810(2) &4.03(2) & 60.7/49  \\   
95437-01-12-06  & 31.2838&  31.3888 &5.9&$3.46_{-0.05 }^{+0.02 }  $&$9.61_{-0.10  }^{+0.04}$&  $0.90_{-0.05 }^{+0.02 }$  & 0.821(2) &4.05(2) & 70.0/49  \\   
95437-01-13-00  & 32.3427&  32.4471 &4.8&$3.46_{-0.09 }^{+0.03 }  $&$9.58_{-0.30  }^{+0.04}$&  $0.90_{-0.07 }^{+0.02 }$  & 0.808(2) &3.98(2) & 54.6/49  \\   
95437-01-13-01  & 33.3886&  33.4799 &4.9&$3.47_{-0.07 }^{+0.02 }  $&$9.51_{-0.10  }^{+0.04}$&  $0.92_{-0.06 }^{+0.02 }$  & 0.801(2) &3.94(2) & 80.6/49  \\   
95437-01-13-02  & 34.3547&  34.4604 &5.4&$3.44_{-0.1  }^{+0.04 }  $&$9.74_{-0.10  }^{+0.07}$&  $0.90_{-0.04 }^{+0.02 }$  & 0.772(1) &3.80(2) & 69.4/49  \\   
95437-01-13-03  & 35.3356&  35.4578 &6.0&$3.46_{-0.05 }^{+0.03 }  $&$9.76_{-0.10  }^{+0.06}$&  $0.89_{-0.04 }^{+0.01 }$  & 0.751(2) &3.70(2) & 78.3/49  \\   
95437-01-13-04  & 36.1199&  36.2499 &5.9&$3.41_{-0.02 }^{+0.06 }  $&$9.93_{-0.04 }^{+0.05}$ &  $0.88_{-0.02 }^{+0.06 }$  & 0.749(1) &3.69(2) & 71.2/49  \\    
95437-01-13-05  & 37.1653&  37.2903 &5.9&$3.50_{-0.03 }^{+0.02 }  $&$9.44_{-0.06 }^{+0.02}$ &  $0.90_{-0.05 }^{+0.02 }$  & 0.690(1) &3.41(2) & 80.7/49  \\    
95437-01-13-06  & 38.2315&  38.3153 &4.3&$3.46_{-0.04 }^{+0.02 }  $&$9.80_{-0.04 }^{+0.07}$ &  $0.89_{-0.03 }^{+0.03 }$  & 0.697(1) &3.43(2) & 75.5/49  \\    
95437-01-14-00  & 39.2564&  39.3715 &4.2&$3.44_{-0.02 }^{+0.02 }  $&$9.82_{-0.02 }^{+0.02}$ &  $0.866_{-0.07 }^{+0.005}$ & 0.654(1) &3.24(2) & 75.0/49  \\

\end{longtable}

\begin{longtable}{lccccccccccc}

\caption[l]{Best-fitting parameters of the spectral continuum
  evaluated from the combined spectra composed of (quasi)simultaneous
  observations performed by XRT on-board \textsl{Swift} (1--10 keV)
  and by PCA on-board \textsl{RXTE} (2.5--30 keV). Columns are:
  Observation ID; Start time of the observation (MJD since $T_0$ = MJD
  55479.81229); End time of the observation (MJD since $T_0$);
  Exposure (ks); $N_{\mbox{H}}$ ($10^{22}$ cm$^{-2}$); $kT_e$ (keV);
  $\tau$; $kT_{S}$ (keV); XRT/PCA normalization; Unabsorbed flux in
  the 2.5--25 keV (PCA) and in the 1-10 keV (XRT) band ($10^{-8}$ erg
  cm$^{-2}$ s$^{-1}$); Bolometric luminosity calculated by
  extrapolating the best-fitting model to the 0.05--150 keV energy
  band ($d_{5.9}^2\:10^{37}$ erg cm$^{-2}$ s$^{-1}$).  The quoted
  errors are evaluated at the 90\% confidence level on each parameter
  ($\Delta\chi^2=2.706$).\label{tab:swxtespectra}}

\\ 
\endfirsthead 
\multicolumn{12}{c}

{{\tablename\	\thetable{}	--	continued	from	previous	page}}	\\				

\hline \hline
 {} & T$_{start}$ (MJD) & T$_{end}$ (MJD) &
Expo (ks) & $N_{\mbox{H}}$ ($10^{22}$ cm$^{-2}$)  & kT$_e$ (keV) & $\tau$ & kT$_{S}$ (keV) & XRT/PCA &
F($10^{-8}$ erg cm$^2$ s$^{-1}$) & L ($d_{5.9}^2\:10^{37}$ erg
s$^{-1}$)&$\chi^2$/dof \\

\hline
\hline						
\endhead																											
\hline	
\multicolumn{12}{c}{{Continued	on	next	page}}	\\
\endfoot
\hline
\endlastfoot
\hline		
											
 ObsID & T$_{start}$ (MJD) & T$_{end}$ (MJD) & Expo (ks) &  $N_{\mbox{H}}$ ($10^{22}$ cm$^{-2}$)  & kT$_e$ (keV) & $\tau$ & kT$_{S}$ (keV) & XRT/PCA &
F($10^{-8}$ erg cm$^2$ s$^{-1}$) & L ($d_{5.9}^2\:10^{37}$ erg s$^{-1}$)& $\chi^2$/dof \\

\hline \hline							

95437-01-03-03(XTE)& 4.9432 &5.1166 &8.0&1.24(4)& $3.30_{-0.03}^{+0.04}$& $9.59_{-0.06}^{+0.04}$&      0.8(2)& $\cdots$   & 1.204(3) &    6.10(2)  &   692/655 \\ 
\vspace{0.2cm}  
00031841003(SW)   & 4.9577 &4.9697 &1.0&   $\cdots$     &  $\cdots$          &        $\cdots$                &   $\cdots$       & 0.940(6)&  1.09(1)  &           &             \\        
95437-01-04-00(XTE)& 5.6019 &5.7049 &3.5&1.24(3)& $3.16_{-0.04}^{+0.05}$& $9.18_{-0.07}^{+0.05}$&      0.8(2)& $\cdots$   &  1.41(4)  &    7.14(2)  &   841/701   \\
\vspace{0.2cm}  
00031841004(SW)   & 5.5469 &5.5611 &1.2&   $\cdots$       &    $\cdots$      &        $\cdots$                &   $\cdots$       & 0.93(1) & 1.32(1)  &           &              \\      
95437-01-07-01(XTE)& 8.9969 &9.0577 &2.3&1.32(4)& $2.80_{-0.07}^{+0.08}$& $7.52_{-0.10}^{+0.08}$&      0.9(2)& $\cdots$   &   1.340(4) &    7.07(3)   &  1024/646 \\
\vspace{0.2cm}  
00031841007(SW)   & 9.02443&9.0336 &0.8&    $\cdots$      &   $\cdots$       &     $\cdots$                   &  $\cdots$        & 0.92(1) &  1.44(1)  &           &             \\      
95437-01-08-00(XTE)& 9.71673&9.84933&5.2&1.16(4)& $2.98_{-0.06}^{+0.08}$& $7.69_{-0.09}^{+0.08}$&      0.9(2)& $\cdots$   &  1.344(4) &    6.96(3)   &  748/652  \\
\vspace{0.2cm}  
00031841008(SW)   & 9.84174&9.85424&1.1&     $\cdots$    &       $\cdots$    &         $\cdots$               & $\cdots$         & 0.96(1) & 1.38(1)  &            &            \\       
95437-01-10-05(XTE)& 16.5797&16.7062&4.8&1.5(1)& $2.89(5)$       &      $10.8(2)$       &   0.8(7)& $\cdots$   & 1.125(7) &    5.61(2)   &   515/358  \\
\vspace{0.2cm}  
00031841013(SW)   & 16.5866&16.5898&0.3&     $\cdots$    &   $\cdots$        &    $\cdots$                    &  $\cdots$        & 0.96(1) &  1.02(2)  &           &             \\      
95437-01-11-01(XTE)& 19.5358&19.6465&4.2&1.37(7)& $3.09_{-0.04}^{+0.06}$& $10.3(1)$        &      0.9(4)&  $\cdots$   & 1.004(4) &    4.90(3)   &   742/614  \\
\vspace{0.2cm}  
00031841016(SW)   & 19.5290&19.5416&1.1&     $\cdots$     &     $\cdots$     &       $\cdots$                 &    $\cdots$       & 0.93(1) & 0.89(1)  &           &             \\      
95437-01-11-02(XTE)& 20.5158&20.6291&4.3&1.10(5)&   $3.21(4)$       &      $10.2_{-0.09}^{+0.08}$&  0.9(4)&  $\cdots$   & 0.991(3) &    4.90(3)   &  579/618  \\
00437466000(SW)   & 20.6155&20.6276&1.0&      $\cdots$    &   $\cdots$       &        $\cdots$                &     $\cdots$      & 0.927(7) &  0.86(1)  &         &\\

\end{longtable}

\begin{longtable}{lcccccccccc}

\caption[l]{Best-fitting parameters of the spectral continuum
  evaluated by combining the 1--10 keV spectrum observed by XRT
  on-board {\it Swift} starting on MJD 55479.801, with the 4.5--25 keV
  JEM-X spectrum and the 20--150 keV ISGRI spectrum accumulated during
  the \textsl{INTEGRAL} revolution No. 975 (MJD 55479.365 --
  55479.548). This group of observations is labeled in text as
  Obs. [A]. Columns are: Observation ID or satellite revolution; Start
  time of the observation (MJD since $T_0$ = MJD 55479.81229); End
  time of the observation (MJD since $T_0$); Exposure (ks);
  $N_{\mbox{H}}$ ($10^{22}$ cm$^{-2}$); $kT_e$ (keV); $\tau$; $kT_{S}$
  (keV); Unabsorbed flux in the 1--10 keV (XRT), 4.5--20 keV (JEM-X)
  or in the 20-150 keV (ISGRI) bands ($10^{-8}$ erg cm$^{-2}$
  s$^{-1}$); Normalisation of JEM-X and of the ISGRI spectra with
  respect to XRT; Reduced chi squared of the fit.  The quoted errors
  are evaluated at the 90\% confidence level on each 
  parameter ($\Delta\chi^2=2.706$).\label{tab:obsAspectra} }

\\ \endfirsthead

 ObsID (Sat. rev.)  & T$_{start}$ (MJD) & T$_{end}$ (MJD) & Expo (ks)
 &$N_{\mbox{H}}$ ($10^{22}$ cm$^{-2}$)& kT$_e$ (keV) & $\tau$ &
 kT$_{S}$ (keV) & F & (JEM-X - ISGRI)/XRT &$\chi^2$/dof \\

\hline																											
\hline

00031841002 & -0.0103 & 0.0127 & 2.0 & 1.1(2) & $20_{-5}^{+20}$ & $4.4_{-0.4}^{+1.4}$ & 0.7(1) & 0.053(6) & $\cdots$ & 109/109 \\
975(JEMX)   & -0.4471 & -0.2640 & 8.7 & $\cdots$ &$\cdots$ & $\cdots$ & $\cdots$ & 0.05(1) & $0.68_{-0.15}^{+0.17}$ & $\cdots$ \\
975(ISGRI)  & -0.4477 & -0.2821 & 8.2 & $\cdots$ &$\cdots$ & $\cdots$ & $\cdots$ & 0.08(1) & $\cdots$ &$\cdots$ \\

\end{longtable}

\begin{longtable}{lcccccccccc}

\caption[l]{Best-fitting parameters of the spectral continuum
  evaluated from the combined 4.5--150 keV spectra observed by JEM-X
  and ISGRI on-board \textsl{INTEGRAL}. Columns are: Satellite
  revolution; Start time of the observation (MJD since $T_0$ = MJD
  55479.81229); End time of the observation (MJD since $T_0$);
  Exposure (ks); $kT_e$ (keV); $\tau$; $kT_{S}$ (keV); Unabsorbed
  flux in the 4.5--20 keV (JEM-X) or in the 20-150 keV (ISGRI) band
  ($10^{-8}$ erg cm$^{-2}$ s$^{-1}$); ISGRI ``soft'' unabsorbed flux
  in the 20--50 keV band ($10^{-8}$ erg cm$^{-2}$ s$^{-1}$); ISGRI
  ``hard' unabsorbed flux in the 50--150 keV band ($10^{-8}$ erg
  cm$^{-2}$ s$^{-1}$); Reduced chi squared of the fit. The quoted
  errors are evaluated at the 90\% confidence level on each parameter
  ($\Delta\chi^2=2.706$).\label{tab:igrspectra} }

\\ \endfirsthead \multicolumn{11}{c}

{{\tablename\	\thetable{}	--	continued	from	previous	page}}	\\				

\hline \hline

 {} & T$_{start}$ (MJD) & T$_{end}$ (MJD) & Expo (ks) & kT$_e$ (keV)
& $\tau$ & kT$_{S}$ (keV) & F($10^{-8}$ erg cm$^2$ s$^{-1}$) &
F$_S$($10^{-8}$ erg cm$^2$ s$^{-1}$) & F$_H$($10^{-8}$ erg cm$^2$
s$^{-1}$)&$\chi^2$/dof \\

\hline																											
\hline

\endhead																											
\hline	\multicolumn{10}{c}{{Continued	on	next	page}}	\\																					
\endfoot																											
\hline																											
\endlastfoot																											
\hline		
																	
 {} & T$_{start}$ (MJD) & T$_{end}$ (MJD) & Expo (ks) & kT$_e$ (keV)
& $\tau$ & kT$_{S}$ (keV) & F($10^{-8}$ erg cm$^2$ s$^{-1}$) &
F$_S$($10^{-8}$ erg cm$^2$ s$^{-1}$) & F$_H$($10^{-8}$ erg cm$^2$
s$^{-1}$)&$\chi^2$/dof \\

\hline																													
\hline

975(JEMX) & -0.4471 & -0.2640 & 8.7 & $15_{-3}^{+7}$     & $6.1_{-0.6}^{+0.9}$ &  $<4.8$    & 0.04(1) &          &         & 10.4/11 \\
\vspace{0.2cm}
975(ISGRI)& -0.4477 & -0.2821 & 8.2 & $\cdots$          & $\cdots$           & $\cdots$   & 0.08(1) & 0.048(4) & 0.04(1) &          \\

976(JEMX) & 0.2889  & 0.4611  &10.0 & $35_{-12}^{+120}$  & $2.2_{-0.3}^{+2.7}$ &  $<1.2$    & 0.12(1) &          &         & 17.1/14 \\
\vspace{0.2cm}
976(ISGRI)& 0.2896  & 0.4609  & 8.4 & $\cdots$          & $\cdots$           & $\cdots$   & 0.11(1) & 0.063(4) & 0.05(1) \\

978(JEMX) & 6.4121  & 6.5878  &11.6 & $3.6_{-0.4}^{+0.7}$& $6.3_{-0.7}^{+0.5}$ &  1.1(1)    & 1.03(1) &  & & 45.4/47 \\
\vspace{0.2cm}
978(ISGRI)& 6.4127  & 6.5876  & 8.7 & $\cdots$          & $\cdots$           & $\cdots$  & 0.038(4) & 0.038(4) & $<5\times10^{-4}$\\

979(JEMX) & 9.0071  & 9.1403  & 4.4 & $>2.1$            & $>5.4$             & $1.3_{-0.3}^{+0.2}$ & 1.00(4) & & & 20.9/44 \\
\vspace{0.2cm}
979(ISGRI)& 9.0078  & 9.1388  & 6.3 & $\cdots$          & $\cdots$           & $\cdots$  &          0.026(6) & 0.025(5) & $<1\times10^{-3}$ \\

980(JEMX) & 11.9746 & 12.1288 & 5.2 & $>1.0$            & $>4.0$             & $1.5_{-0.3}^{+0.1}$ & 0.75(3) & & & 21.9/45 \\
\vspace{0.2cm}
980(ISGRI)& 11.9746 & 12.1288 & 7.5 & $\cdots$          & $\cdots$           & $\cdots$  &    0.021(5) & 0.021(5) & $<2\times10^{-3}$\\

981(ISGRI)\footnote{The quoted fluxes are evaluated by fitting the
  ISGRI points with a power law model.}
& 14.9667 & 17.6446 & 121.4 &$\cdots$ & $\cdots$ & $\cdots$ & 0.023(5) & 0.023(5) & $<3\times10^{-3}$ & 3.93/1\\
\end{longtable}

\end{landscape}

\end{center}																				

\end{document}